\documentclass{aa}
\usepackage[dvips]{graphicx}
\usepackage{natbib}
\usepackage{txfonts}
\usepackage{color}

\begin{document}
   \title{Radiative hydrodynamics simulations of red supergiant stars:
     II. simulations of convection on Betelgeuse match
interferometric observations}
\titlerunning{Convection on Betelgeuse}
%   \subtitle{I. Overviewing the $\kappa$-mechanism}

   \author{A. Chiavassa
          \inst{1,2}
          \and
          X. Haubois \inst{3}
          %\fnmsep\thanks{Just to show the usage
%          of the elements in the author field}
          \and
          J. S. Young \inst{4}
          \and
          B. Plez\inst{2}
          \and
          E. Josselin\inst{2}
          \and
          G. Perrin \inst{3}
          \and
          B. Freytag\inst{5,6}
          }

   \offprints{A. Chiavassa}

    \institute{Max-Planck-Institut f\"{u}r Astrophysik, Karl-Schwarzschild-Str. 1, Postfach 1317, DÐ85741 Garching b. M\"{u}nchen, Germany\\
              \email{chiavass@mpa-garching.mpg.de}
            \and
            GRAAL, Universit\'{e} de Montpellier II - IPM, CNRS, Place Eug\'{e}ne Bataillon
	34095 Montpellier Cedex 05, France
         \and
         Observatoire de Paris, LESIA, UMR 8109, 92190 Meudon, France
         \and
         Astrophysics Group, Cavendish Laboratory, JJ Thomson Avenue, Cambridge CB3 0HE
         \and
             Centre de Recherche Astrophysique de Lyon,
          UMR 5574: CNRS, Universit\'e de Lyon,
          \'Ecole Normale Sup\'erieure de Lyon,
          46 all\'ee d'Italie, F-69364 Lyon Cedex 07, France
            \and
          Department of Physics and Astronomy,
          Division of Astronomy and Space Physics,
          Uppsala University,
          Box 515, S-751~ 20 Uppsala,
          Sweden
             }

   \date{Received; accepted }

  \abstract
  % context heading (optional), leave it empty if necessary
 {The red supergiant (RSG) Betelgeuse is an irregular variable
   star. Convection may play an important role in understanding this
   variability. Interferometric observations can be interpreted using
   sophisticated simulations of stellar convection.}
 {We compare the visibility curves and
 closure phases obtained from our 3D simulation of RSG convection with
 CO5BOLD to various interferometric observations of Betelgeuse from
 the optical to the H band in order to characterize and measure the
 convection pattern on this star.}
 {We use 3D radiative-hydrodynamics
   (RHD) simulation to compute intensity maps in different filters and
   we thus derive interferometric observables using the
   post-processing radiative transfer code OPTIM3D. The synthetic
   visibility curves and closure phases are compared to observations.}
 {We provide a robust detection of the
     granulation pattern on the surface of Betelgeuse in the optical
     and in the H band based on excellent fits to the observed
     visibility points and closure phases. Moreover, we determine that
     the Betelgeuse surface in the H band is covered by small to
     medium scale (5--15 mas) convection-related surface structures
     and a large ($\approx$30 mas) convective cell. In this spectral
     region, H$_2$O molecules are the main absorbers and contribute to
     the small structures and to the position of the first null of the
     visibility curve (i.e.\ the apparent stellar radius).}
   {}


   \maketitle
   
%
%________________________________________________________________

\section{Introduction}

Betelgeuse is a red supergiant star (Betelgeuse, HD~39801,
M1--2Ia--Ibe) and is one of the brightest stars in the optical and
near infrared. This star exhibits variations in integrated brightness,
surface features, and the depths, shapes, and Doppler shifts of its
spectral lines. There is a backlog of visual-wavelength observations
of its brightness which covers almost a hundred years. The irregular
fluctuations of its light curve are clearly aperiodic and rather
resemble a series of outbursts. \cite{2006MNRAS.372.1721K} studied the
variability of different red supergiant (RSG) stars including
Betelgeuse and they found a strong noise component in the photometric
variability, probably caused by the large convection cells.  In addition to
this, the spectral line variations have been analyzed by several
authors, who inferred the presence of large granules and
high convective velocities
\citep{2007A&A...469..671J,2008AJ....135.1450G}.
\citeauthor{2008AJ....135.1450G} also found line bisectors that
predominantly have reversed C-shapes, and line shape variations
occurring at the 1 km\,s$^{-1}$ level that have no obvious connection to
their shifts in wavelength.

The position of Betelgeuse on the H-R diagram is highly uncertain, due
to uncertainty in its effective
temperature. \cite{2005ApJ...628..973L} used one-dimensional MARCS
models \citep{2003ASPC..288..331G,1975A&A....42..407G} to fit the
incredibly rich TiO molecular bands in the optical region of the
spectrum for several RSGs. They found an effective temperature of
3650~K for Betelgeuse. Despite the fact that they obtained a good
agreement with the evolutionary tracks, problems remain. There is a
mismatch in the IR colors that could be due to atmospheric
temperature inhomogeneities characteristic of convection
\citep{2006ApJ...645.1102L}. Also the distance of Betelgeuse has large uncertainities because of errors related to the positional movement of the stellar photocenter. \cite{2008AJ....135.1430H} derived a distance of (197$\pm$45 pc) using high spatial resolution, multiwavelength, VLA radio positions combined with Hipparcos Catalogue Intermediate Astrometric Data.

Betelgeuse is one of the best studied RSGs in term of multi-wavelength
imaging because of its large luminosity and angular diameter. The
existence of hot spots on its surface has been proposed to explain
numerous interferometric observations with WHT and COAST
\citep{1990MNRAS.245P...7B, 1992MNRAS.257..369W, 1997MNRAS.285..529T,
  1997MNRAS.291..819W, 2000MNRAS.315..635Y, 2004young} that have
detected time-variable inhomogeneities in the brightness
distribution. These authors fitted the visibility and closure phase
data with a circular limb-darkened disk and zero to three spots. A
large spot has been also detected by \cite{1998AJ....116.2501U} with
HST. The non-spherical shape of Betelgeuse was also detected by
\cite{2007ApJ...670L..21T} in the
mid-infrared. \cite{2009A&A...508..923H} published a reconstructed
image of Betelgeuse in the H band with two spots using the same data
as presented in this work. \cite{2009A&A...504..115K} resolved
Betelgeuse using diffraction-limited adaptive optics in the
near-infrared and found an asymmetric envelope around the star with a
bright plume extending in the southwestern region. They claimed the
plume was either due to the presence of a convective hot spot or
was caused by stellar rotation. \cite{2009A&A...503..183O}
presented VLTI/AMBER observations of Betelgeuse at high spectral
resolution and spatially resolved CO gas motions. They claimed that
these motions were related to convective motions in the upper
atmosphere or to intermittent mass ejections in clumps or arcs.

Radiation hydrodynamics (RHD) simulations of red supergiant stars are
available \citep{2002AN....323..213F} to interpret past and future
observations.  \cite[hereafter Paper~I]{2009A&A...506.1351C} used these
simulations to explore the impact of the granulation pattern on
observed visibility curves and closure phases and detected a
granulation pattern on Betelgeuse in the K band by fitting the
existing interferometric data of \cite{2004A&A...418..675P}.

This paper is the second in the series aimed at exploring the
convection in RSGs. The main purpose is to compare RHD simulations to
high-angular resolution observations of Betelgeuse covering a wide
spectral range from the optical region to the near-infrared H band, in
order to confirm the presence of convective cells on its surface.

\section{3D radiation-hydrodynamics simulations and post-processing radiative transfer}\label{paramSect}

We employed numerical simulations obtained using CO$^5$BOLD
\citep{2002AN....323..213F, Freytag2003SPIE.4838..348F,
  Freytag2008A&A...483..571F} and in particular the model st35gm03n07
that has been deeply analyzed in Paper~I. The model has 12 $M_{\odot}$,
a numerical resolution of 235$^3$ grid points with a step of
8.6~$R_{\odot}$, an average luminosity over spherical shells and over
time of $L$=93000$\pm$1300~$L_{\odot}$, an effective temperature of
$T_{\rm{eff}}$=3490$\pm$13~K, a radius of $R$=832$\pm$0.7~$R_{\odot}$,
and surface gravity log($g$)=-0.337$\pm$0.001. This is our ``best''
RHD simulation so far because it has stellar parameters closest to
Betelgeuse ($T_{\rm{eff}}=3640$~K, \citeauthor{2005ApJ...628..973L},
\citeyear{2005ApJ...628..973L}; and log($g$)=-0.3, \citeauthor{2008AJ....135.1430H}, \citeyear{2008AJ....135.1430H}).

We used the 3D pure-LTE radiative transfer code OPTIM3D described in
Paper~I to compute intensity maps from all the suitable snapshots of
the 3D hydrodynamical simulation. The code takes into account the
Doppler shifts caused by the convective motions. The radiation
transfer is calculated in detail using pre-tabulated extinction
coefficients generated with the MARCS code
\citep{2008A&A...486..951G}. These tables are functions of
temperature, density and wavelength, and were computed with the solar
composition of \citet{2006CoAst.147...76A}. The tables include the
same extensive atomic and molecular data as the MARCS models. They
were constructed with no micro-turbulence broadening and the
temperature and density distributions are optimized to cover the values
encountered in the outer layers of the RHD simulations.

\section{Observations}

The data presented in this work have been taken by two independent
groups with different telescopes and they cover a large wavelength
range from the optical to the near infrared. The log of the observations is
reported in Table~\ref{tab:Logob}

\begin{table}
\begin{minipage}[t]{\columnwidth}
\caption{Log of the observations.}
\label{tab:Logob}
\centering
\begin{tabular}{ccc}
\hline\hline
Date & Telescope & Filter (central $\lambda$) \\ 
\hline
October 7, 2005 &  IOTA    & IONIC - 16000  
\\
October 8, 2005 &  IOTA     & IONIC - 16000 
\\
October 10, 2005&  IOTA   &  IONIC - 16000 
\\
October 11, 2005&  IOTA    & IONIC - 16000 
\\
October 12, 2005&  IOTA    & IONIC  - 16000 
\\
October 16, 2005&  IOTA    &  IONIC - 16000 
\\
\hline
October 21, 1997 & COAST & 12900 \AA  \\
October 24, 1997 & COAST & 9050 \AA  \\
October 31, 1997 & COAST & 9050 \AA  \\ 
November 11, 1997 & COAST & 12900 \AA  \\ 
November 12, 1997 & COAST & 9050 \AA   \\ 
November 15, 1997 & WHT &  7000 \AA   \\
November 16, 1997 & WHT &  9050  \AA \\
November 21, 1997 & COAST & 9050 \AA \\
\hline
January 29, 2004  & COAST & 7500, 7820, 9050 \AA \\
February 8, 2004 & COAST & 7500, 7820, 9050 \AA \\
February 25, 2004 & COAST &7500, 7820, 9050 \AA \\
February 29, 2004 & COAST &7500, 7820, 9050 \AA \\
March 1, 2004 & COAST & 7500, 7820, 9050 \AA \\
March 2, 2004 & COAST & 7500, 7820, 9050 \AA \\
\hline\hline
\end{tabular}
\end{minipage}
\end{table}

\subsection{Data at 16400 \AA.}

The H band data were acquired with the 3 telescope interferometer IOTA
\citep[Infrared Optical Telescope Array,][]{2003SPIE.4838...45T}
located at Mount Hopkins in Arizona. Light collected by three
apertures (siderostats of 0.45m diameter) was spatially filtered by
single mode fibers to clean the wavefronts, removing high frequency
atmospheric corrugations that affect the fringe contrast. The beams
were then combined with IONIC \citep{2003SPIE.4838.1099B}. This
integrated optics component combines 3 input beams in a pairwise
manner. Fringes were encoded in the time domain using piezo-electric
path modulators, and detected with a near-infrared camera utilizing a
PICNIC detector \citep{2004PASP..116..377P}.

\begin{figure}
   \centering
   	 \begin{tabular}{c}
             \includegraphics[width=1.0\hsize]{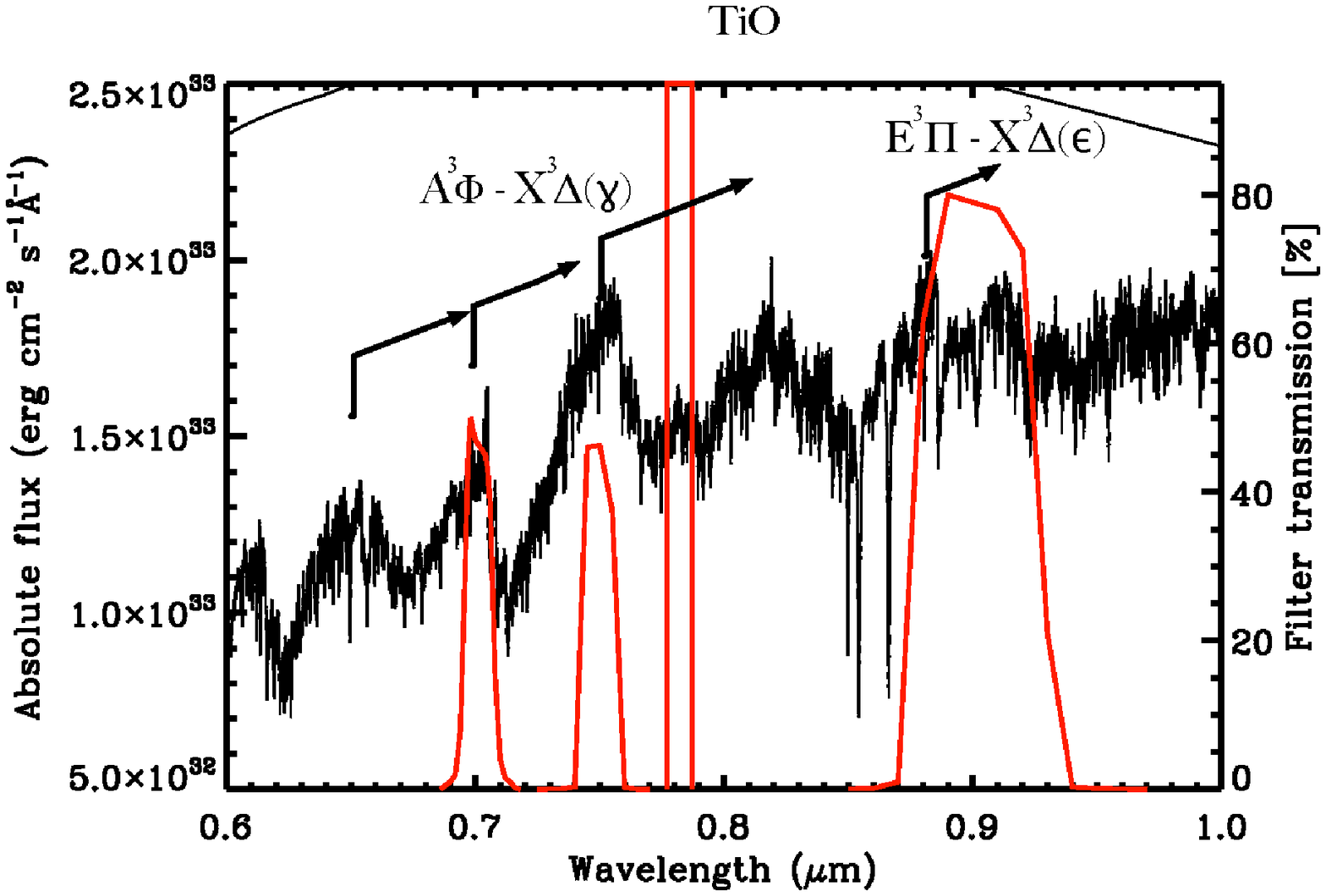}\\
             \includegraphics[width=1.0\hsize]{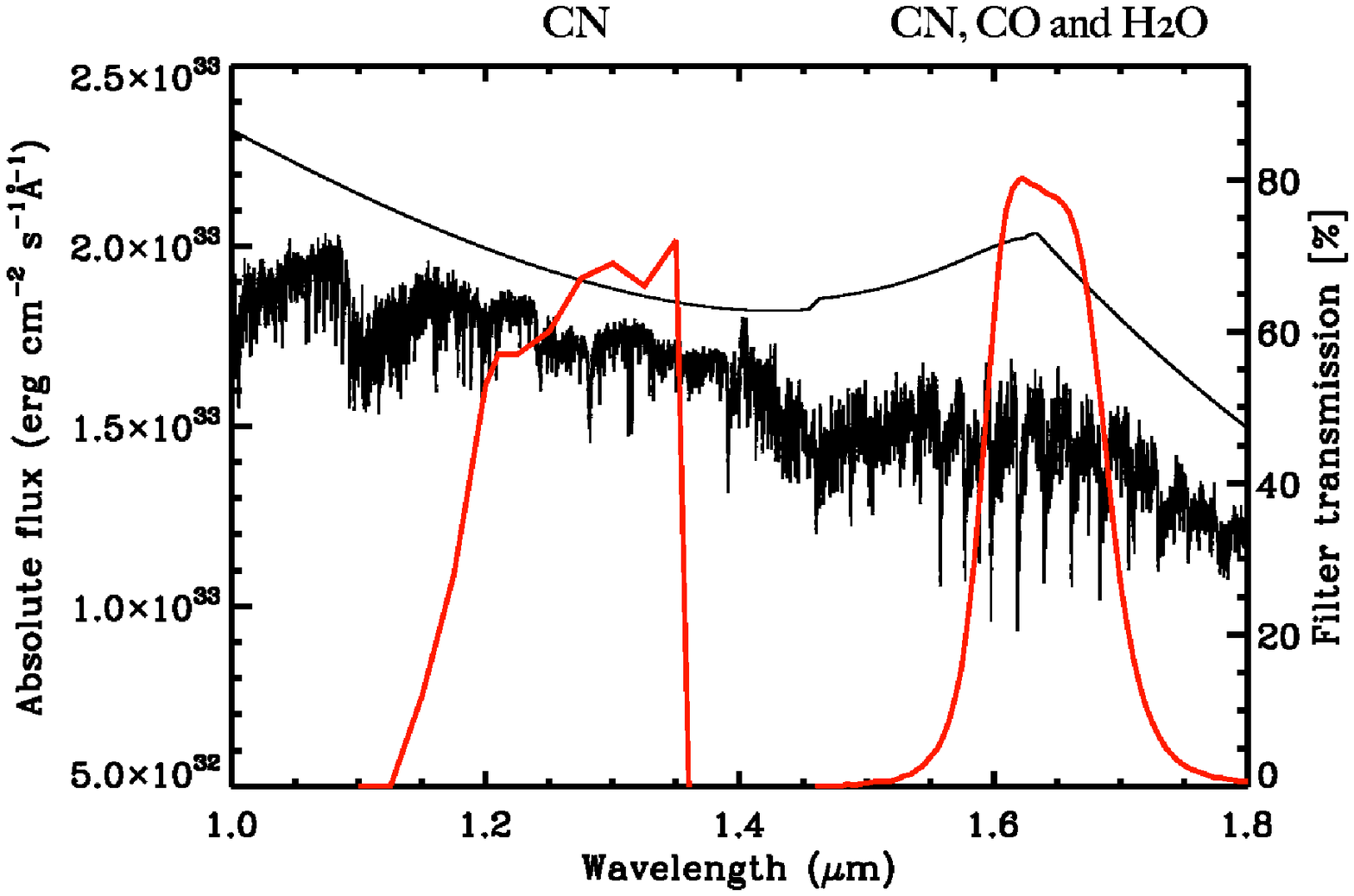}
             \end{tabular}
   \caption{Transmission curves (red) of the bandpass filters used for
  Betelgeuse observations at COAST and WHT: central wavelengths at
  7000, 7500, 7820, 12900 \AA\ and nominal bandwidths (FWHM) of 100,
  130, 50, 500, 1500 \AA\ respectively
  \citep{2004young,2000MNRAS.315..635Y} (the filter curve for the
  narrow bandpass centered at 7820 \AA\ has been lost hence we assumed
  a top-hat filter); and for Betelgeuse observations at IOTA: 16400
  \AA\ with 1000 \AA\ bandwidth \citep{2009A&A...508..923H}. The black
  line is the synthetic spectrum with its continuum computed
  from the RHD simulation snapshot of Fig.~\ref{young_comparison2}; the molecules that
  contribute the most in every filter have been highlighted.}
              \label{filters}%
\end{figure}

Betelgeuse was observed in the H band ($16400\pm1000$ \AA,
Fig.~\ref{filters}) on 6 nights between 7th October and 16th October
2005. Five different configurations of the interferometer telescopes
were used, in order to cover a large range of spatial frequencies
between 12 and 95 arcsec$^{-1}$. To calibrate the instrumental
transfer function, observations of Betelgeuse were interleaved with
observations of a reference (calibrator) star, HD 36167.

Data reduction was carried out using an IDL pipeline
\citep{2004ApJ...602L..57M,2007ApJ...659..626Z}. In order to measure
the closure phase, we took the phase of the complex triple product
(bispectrum, \citeauthor{baldwin86-closure}, \citeyear{baldwin86-closure}). The instrumental closure
phase of IONIC3 drifted less than 1 degree over many hours owing to
the miniature dimensions of the integrated optics component. Both for
the squared visibilities and the closure phase, the random errors were
calculated with the bootstrap technique, in which a statistic is
repeatedly re-estimated by Monte-Carlo sampling the original data with
replacement. Full details of the observations and data reduction can
be found in \cite{2009A&A...508..923H}.

\subsection{Data from 7000 to 12900 \AA.}

For this wavelength range, we used data taken at two different
epochs. The observations carried out in 1997
\citep{2000MNRAS.315..635Y} were acquired with the Cambridge Optical
Aperture Synthesis Telescope (COAST) on baselines up to 8.9 m (with
central wavelengths/bandwidths of 9050/500 and 12900/1500 \AA) and by
non-redundant aperture masking with the William Herschel Telescope
(WHT) on baselines up to 3.7 m (with central
wavelengths/bandwidths of 7000/100 and 9050/500). The observations
carried out in 2004 were obtained with COAST on baselines up to 6.1 m
(bandpasses 7500/130, 7820/50 and 9050/500 \AA). Fig.~\ref{filters}
shows all the filters used.

\subsubsection{COAST data from 1997}

The COAST data were taken during October and November
1997. Observations at 9050 \AA\ were made using the standard
beam-combiner and avalanche photodiode detectors
\cite{1994SPIE.2200..118B}, while 12900 \AA\ observations were
obtained with a separate pupil-plane combiner optimized for JHK bands
\cite{1998SPIE.3350..746Y}.

Observations of Betelgeuse were interleaved with observations of
calibrator stars, either unresolved or of small and known
diameters. If at least three baselines were measurable and the
atmospheric coherence time was sufficiently long, closure phase
measurements were also collected, by recording fringes on three
baselines simultaneously.

Data reduction was carried out using standard methods in which the
power spectrum and bispectrum of the interference fringes were
averaged over each dataset \cite{1997MNRAS.290L..11B}. The resulting
visibilities had formal fractional errors in the range 2--10 $\%$ of
the values and the closure phases had typical uncertainties of
5--10$^{\circ}$. Additional uncertainties of 10--20\%
were added to the visibility amplitudes to accommodate potential
changes in the seeing conditions between observations of the science
target and calibrator stars.

\subsubsection{WHT data from 1997}

The observations with the WHT used the non-redundant aperture masking
method \citep{baldwin86-closure,haniff87-first} and employed a
five-hole linear aperture mask. Filters centred at 7000 \AA\ and 9050
\AA\ were used to select the observing waveband; only the 7000
\AA\ data are presented in this paper. The resulting interference
fringes were imaged onto a CCD and one-dimensional fringe snapshots
were recorded at 12-ms intervals. For each orientation of the mask,
the fringe data were reduced using standard procedures
\citep{haniff87-first,1990MNRAS.245P...7B} to give estimates of the
visibility amplitudes on all 10 interferometer baselines and of the
closure phases on the 10 (linear) triangles of baselines. As for the
COAST measurements, the uncertainties on the visibility amplitudes
were dominated by calibration errors, which in this instance were
unusually large (fractional error $\sim$30\%). On the other hand, the
calibrated closure phase measurements had typical errors of only
1--3$^{\circ}$. The orientation and scale of the detector were
determined by observations of two close visual binaries with
well-determined orbits.

\subsubsection{COAST data from 2004}

The observations taken in 2004 \citep{2004young} were acquired with
COAST using the standard beam combiner and filters centered at 7500,
7820 and 9050 \AA\ with FWHM of 130, 50, and 500 \AA respectively.
The raw interference fringe data were reduced using the same methods
utilised for the 1997 COAST data to obtain a set of estimates of the
visibility amplitude and closure phase for each observing waveband.

\section{Comparison of simulations and observations}

In this section, we compare the synthetic visibility curves and
closure phases to the observations. For this purpose, we used all the
snapshots from the RSG simulation to compute intensity maps with
OPTIM3D. These maps were normalized to the filter transmissions of
Fig.~\ref{filters} as: $\frac{\int I_{\lambda}
  \tau\left(\lambda\right)d\lambda}{\int
  \tau\left(\lambda\right)d\lambda}$ where $I\left(\lambda\right)$ is
the intensity and $\tau\left(\lambda\right)$ is the optical
transmission of the filter at a certain wavelength. Then, for each
intensity map, a discrete Fourier transform ($FT$) was calculated. The
visibility $V$ is defined as the modulus $|z|$ of the complex Fourier
transform $z = x + iy$ (where $x$ is real part of the complex number $z$ and $y$ its imaginary part) normalized to the modulus at the origin of the
frequency plane $|z_0|$, with the phase $\theta$ defined as
$\tan\theta = \Im(z)/\Re(z)$. The closure phase is defined as the
phase of the triple product (or bispectrum) of the complex
visibilities on three baselines, which form a closed loop joining
three stations A, B, and C. If the projection of the baseline AB is
$\left(u_1,v_1\right)$, that for BC is $\left(u_2,v_2\right)$, and
thus $\left(u_1+u_2,v_1+v_2\right)$ for AC, the closure phase is:
\begin{eqnarray*}
\lefteqn{\phi_C (u_1,v_1,u_2,v_2) = }\\
& \arg ( V(u_1,v_1) & \times V(u_2,v_2) \times V^*(u_1+u_2,v_1+v_2) ) .
\end{eqnarray*}
The projected baselines and stations are those of the observations.

Following the method explained in Paper~I, we computed visibility
curves and closure phases for 36 different rotation angles with a step of
5$^\circ$ from all the available intensity maps ($\approx$3.5 years of
stellar time), giving a total of $\approx$2000 synthetic
visibilities and $\approx$2000 synthetic closure phases per filter.

\subsection{Data at 16400 \AA}
\label{ionicsect}

We begin by comparing with the 16400 \AA\ data because this filter is
centered where the H$^{-1}$ continuous opacity minimum
occurs. Consequently, the continuum-forming region is more visible and
the granulation pattern is characterized by large scale granules of
about 400--500 $R_{\odot}$ ($\approx$60$\%$ of the stellar radius)
evolving on a timescale of years (Fig.~4 in Paper~I). On the top of
these cells, there are short-lived (a few months to one year)
small-scale (about 50--100 R$_{\odot}$) structures. The resulting
granulation pattern causes significant fluctuations of the visibility
curves and the signal to be expected in the second, third and fourth
lobes deviates greatly from that predicted by uniform disk (UD) and
limb-darkened disk (LD) models (Fig.~11 in Paper~I). Also the closure
phases show large departures from 0 and $\pm\pi$, the values which
would indicate a point-symmetric brightness distribution.

\begin{figure}[!h]
   \centering
  \begin{tabular}{c}
     \includegraphics[width=1.0\hsize]{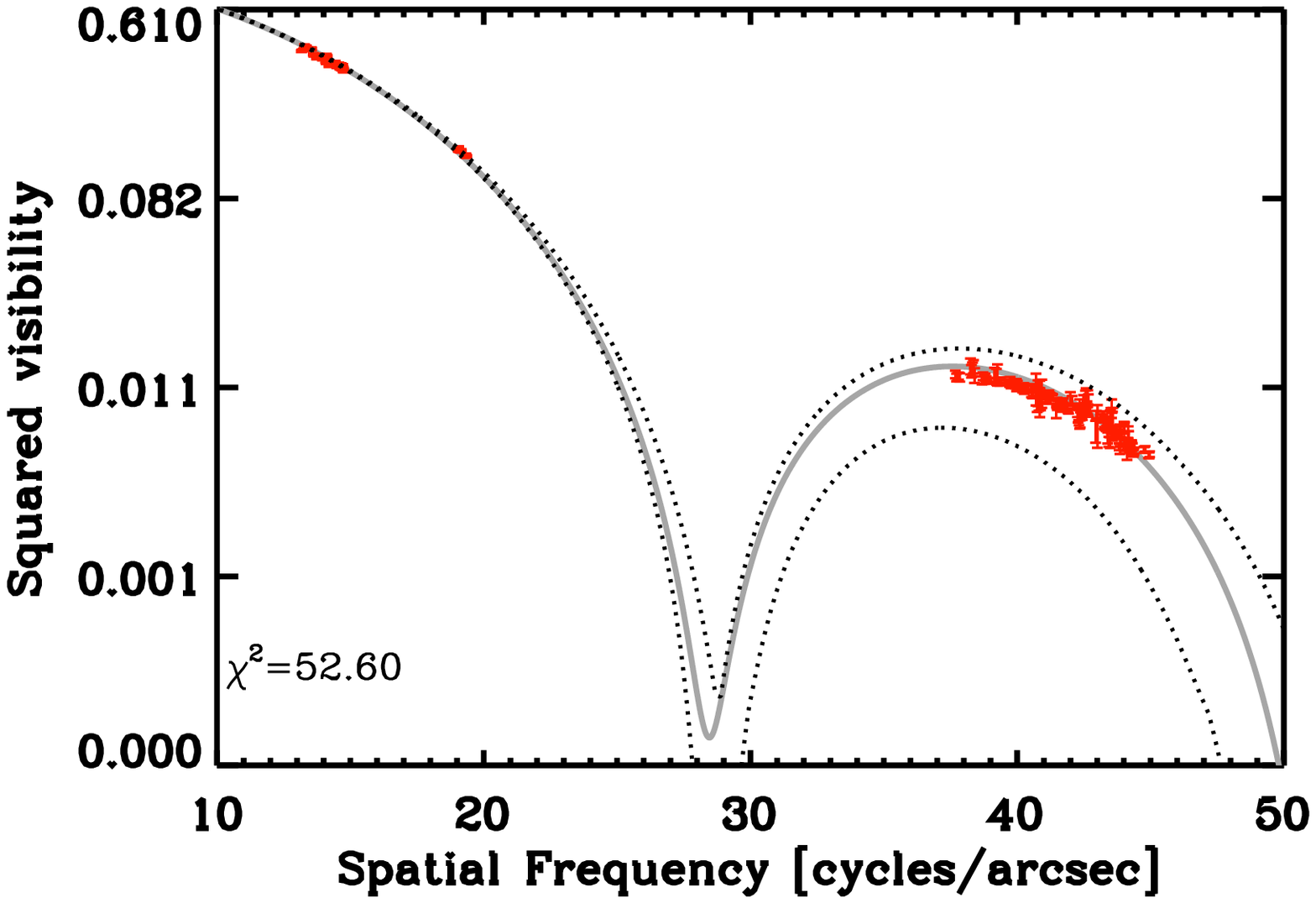}\\
     \includegraphics[width=1.0\hsize]{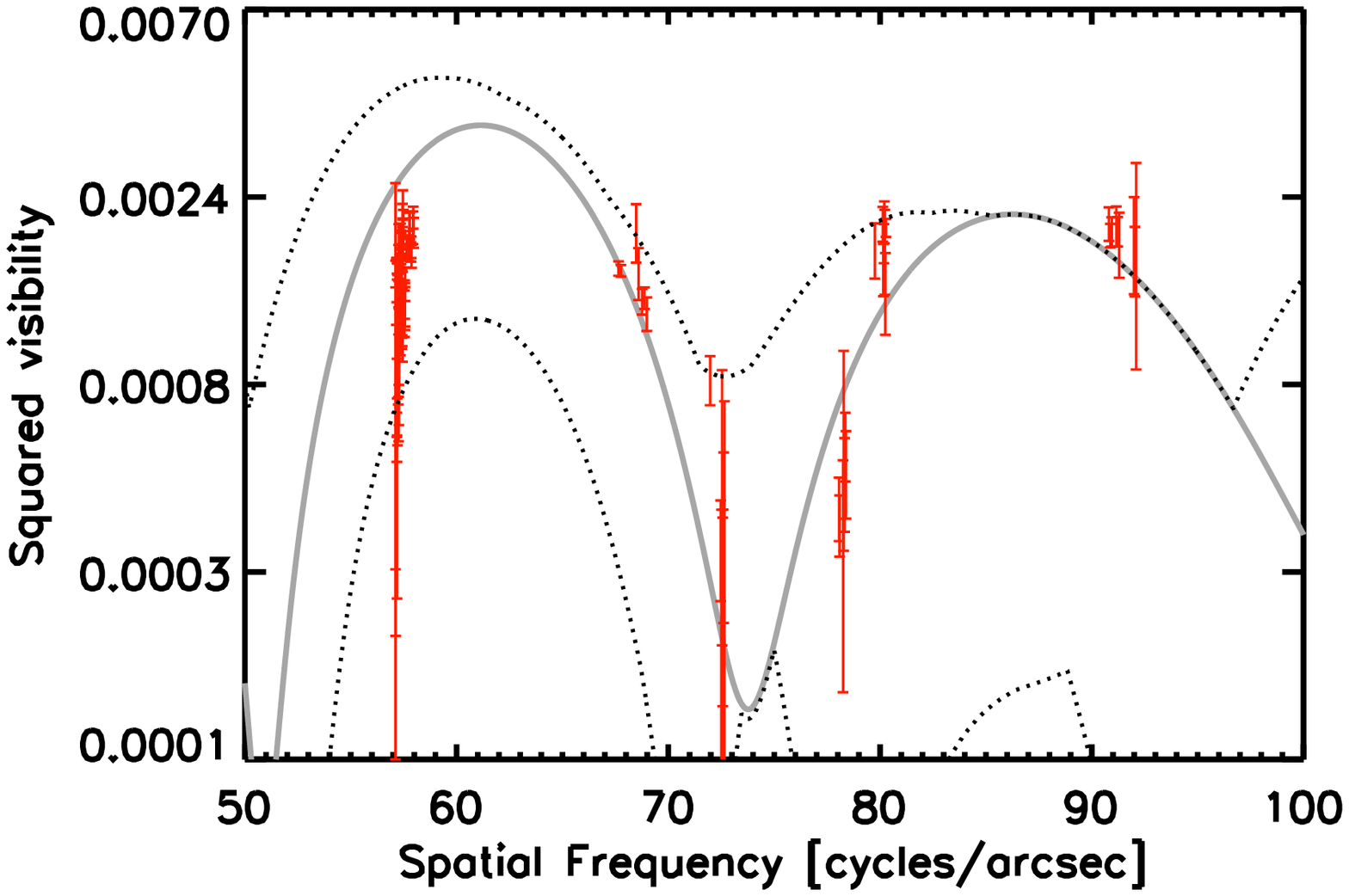}\\
       \includegraphics[width=1.0\hsize]{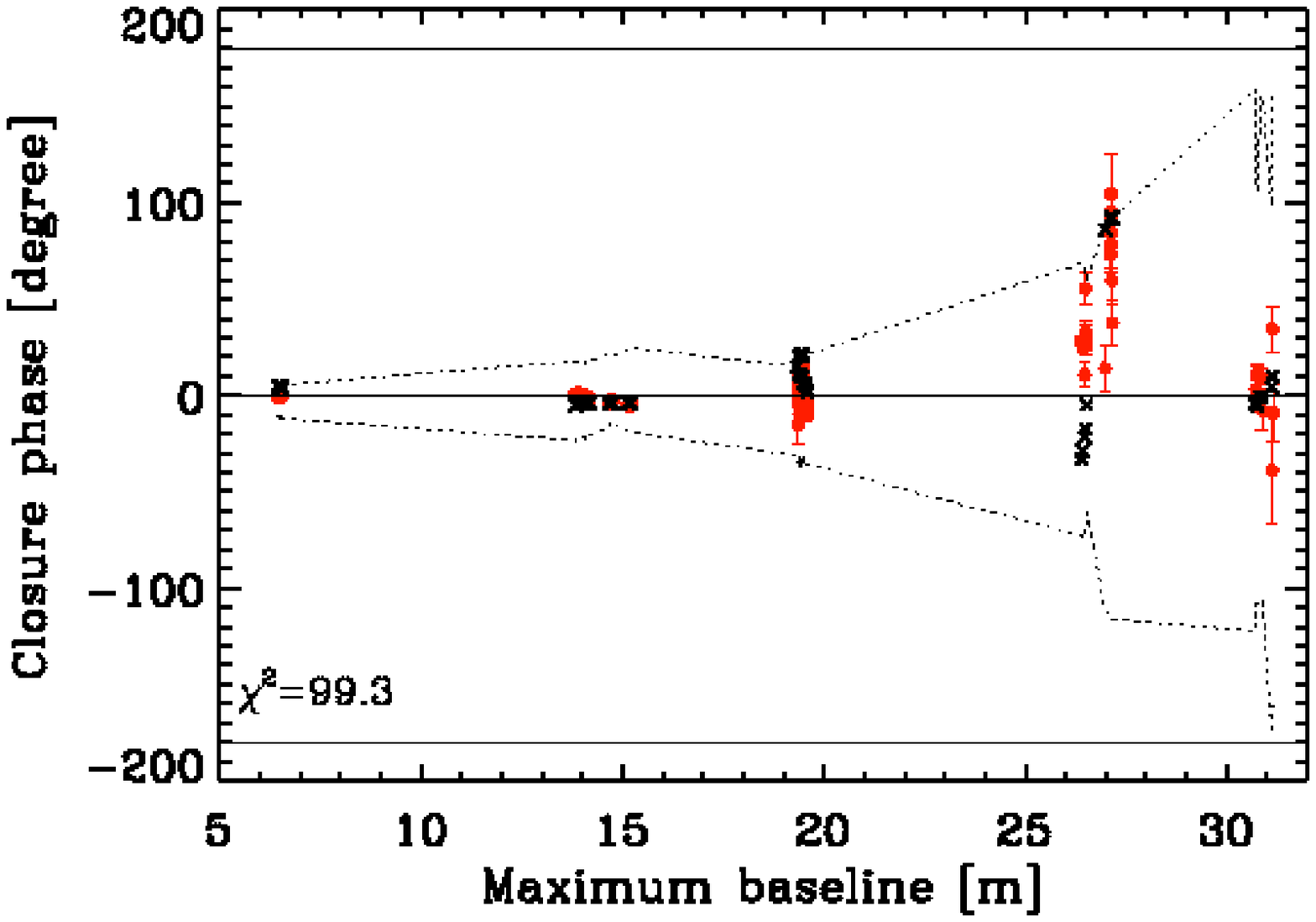}
 \end{tabular}
\caption{\emph{Top and central panels:}
  the best-matching synthetic squared visibility (grey) compared to
  the observations of Betelgeuse
  \citep[][red]{2009A&A...508..923H}. The reduced $\chi^2$ is
  indicated. The dotted lines are the maximum and minimum amplitude of
  the visibility fluctuations as the snapshot is rotated. \emph{Bottom panel:}
  closure phase (in degrees) versus the maximum projected baseline
  length for each baseline triplet. The observations (red dots with
  error bars) are fitted with the same intensity map and rotation
  angle used for the visibilities (black crosses). The axisymmetric
  case is represented by the solid black lines. The dotted lines are
  the maximum and minimum of the closure phase variation as the
  snapshot is rotated.  }
              \label{comparison_1a}%
\end{figure}

Within the large number of computed visibilities and closure phases
for this filter, we found that some match the observation data very
well (Fig.~\ref{comparison_1a}). We selected the best-fitting
snapshot minimizing the function:
\begin{equation}
\chi^2=\frac{1}{N}\left[\sum^{N_{V}}_{i=1}\left(\frac{V_i-M_{V_{i}}}{\sigma_{V_{i}}}\right)^2+\sum^{N_{\phi_C}}_{i=1}\left(\frac{\phi_{C_{i}}-M_{\phi_{C_{i}}}}{\sigma_{\phi_{C_{i}}}}\right)^2\right],
\end{equation}
 where $V_i$ is the observed visibility amplitude data with its
  corresponding error $\sigma_{V_{i}}$, $M_{V_{i}}$ is the synthetic
  visibility amplitude at the same spatial frequency, $\phi_{C_{i}}$ is
  the observed closure phase with corresponding error
  $\sigma_{\phi_{C_{i}}}$, and $M_{\phi_{C_{i}}}$ the synthetic closure
  phase for the observed UV coordinates.  The best matching visibilities
  and closure phases correspond to a particular snapshot and rotation
  angle.

\begin{figure}[!h]
   \centering
\includegraphics[width=0.9\hsize]{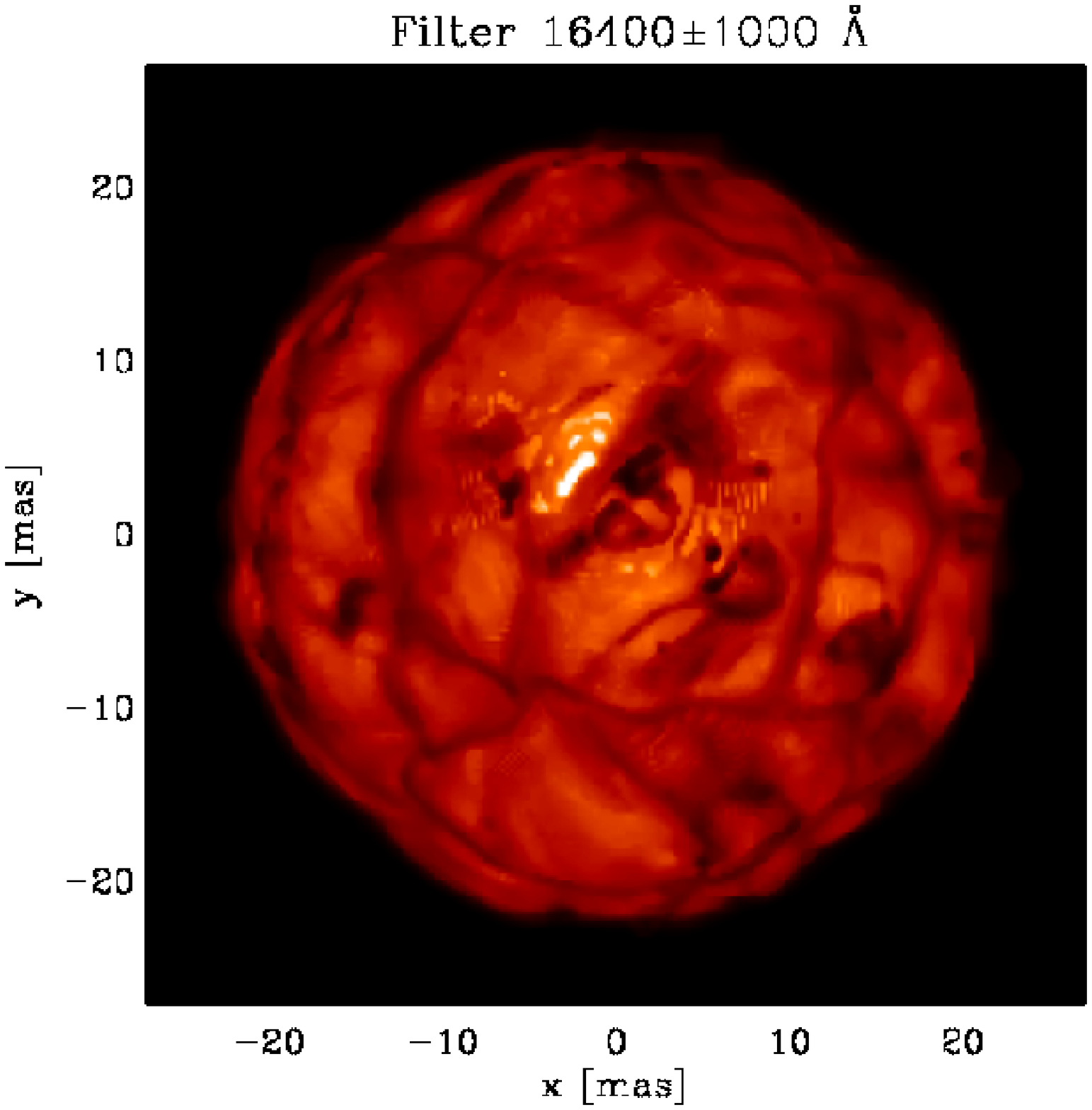}
\caption{Map of the linear intensity in the
  IONIC filter. The range is
  [0;$3.1\times10^5$]\,erg\,cm$^{-2}$\,s$^{-1}$\,{\AA}$^{-1}$. The
  stellar parameters of this snapshot are: $L = 92\,300\,L_\odot$, $R
  = 832.4\,R_\odot$, $T_{\rm eff} = 3486$\,K and $\log(g) =
  -0.34$. The simulation has been scaled to an apparent diameter of
  $\sim$45.1 mas, at a distance of 172.1 pc in order to fit the data
  points in the first lobe.}
\label{comparison_1b}%
\end{figure}

In Fig.~\ref{comparison_1b} the simulation has been scaled to an
apparent diameter of $\sim$45.1 mas in order to fit the data points in
the first lobe, corresponding to a distance of 172.1 pc for the
simulated star. The angular diameter is slightly larger than the limb-darkened
diameter of 44.28$\pm$0.15 mas found by
\cite{2009A&A...508..923H}. Our distance is also in agreement with
\cite{2008AJ....135.1430H}, who reported a distance of
$197\pm45$~pc. Using Harper et al.'s distance and an effective
temperature of 3650 K \citep{2005ApJ...628..973L}, the radius is
$R=890\pm200 R_\odot$, neglecting any uncertainty in $T_{\rm eff}$. On
the other hand, using Harper et al. 2008 distance and the apparent diameter of 45 mas
\citep{2004A&A...418..675P}, the radius is $R=950\pm200 R_\odot$. All
these results match evolutionary tracks by \cite{2003A&A...404..975M}
for an initial mass of between 15 and 25 $M_\odot$. The radius ($R\approx832 R_\odot$, see Section~\ref{paramSect})
and the effective temperature ($T_{\rm eff}\approx3490$ K) of our
3D simulation are smaller because the simulations start with an initial model that has a guessed radius,
a certain envelope mass, a certain potential profile, and a prescribed
luminosity. However, during the run the internal structure relaxes so something
 not to far away from the initial guess (otherwise the numerical grid
is inappropriate). The average final radius is determined once the
simulation is finished. Therefore, since the radius (and the effective
temperature) cannot be tuned, the model is placed at some distance in
order to provide the angular diameter that best matches the
observations. Finally, within the error bars our model radius agrees
with all other data derived using the distance determined by \cite{2008AJ....135.1430H}.\\

Our RHD simulation provides a better
fit than uniform disk and limb-darkened models used by \citeauthor{2009A&A...508..923H} in all lobes of the
visibility function.  The departure from circular symmetry is more
evident at high spatial frequencies (e.g., the fourth lobe) where the
visibility predicted from the parametric model is lower than the
observed data. The small-scale convection-related surface structures
are the cause of this departure and can only be explained by RHD
simulations that are permeated with irregular convection-related
structures of different size.

Also the closure phases display a good agreement with the
simulation indicating that a possible solution to the distribution of
the inhomogeneities on the surface of Betelgeuse is the intensity map
of Fig.~\ref{comparison_1b} (though the reconstructed images found by
\citet{2009A&A...508..923H} are more probable).

\begin{figure*}[!htb]
   \centering
  \begin{tabular}{cc}
\includegraphics[width=0.32\hsize]{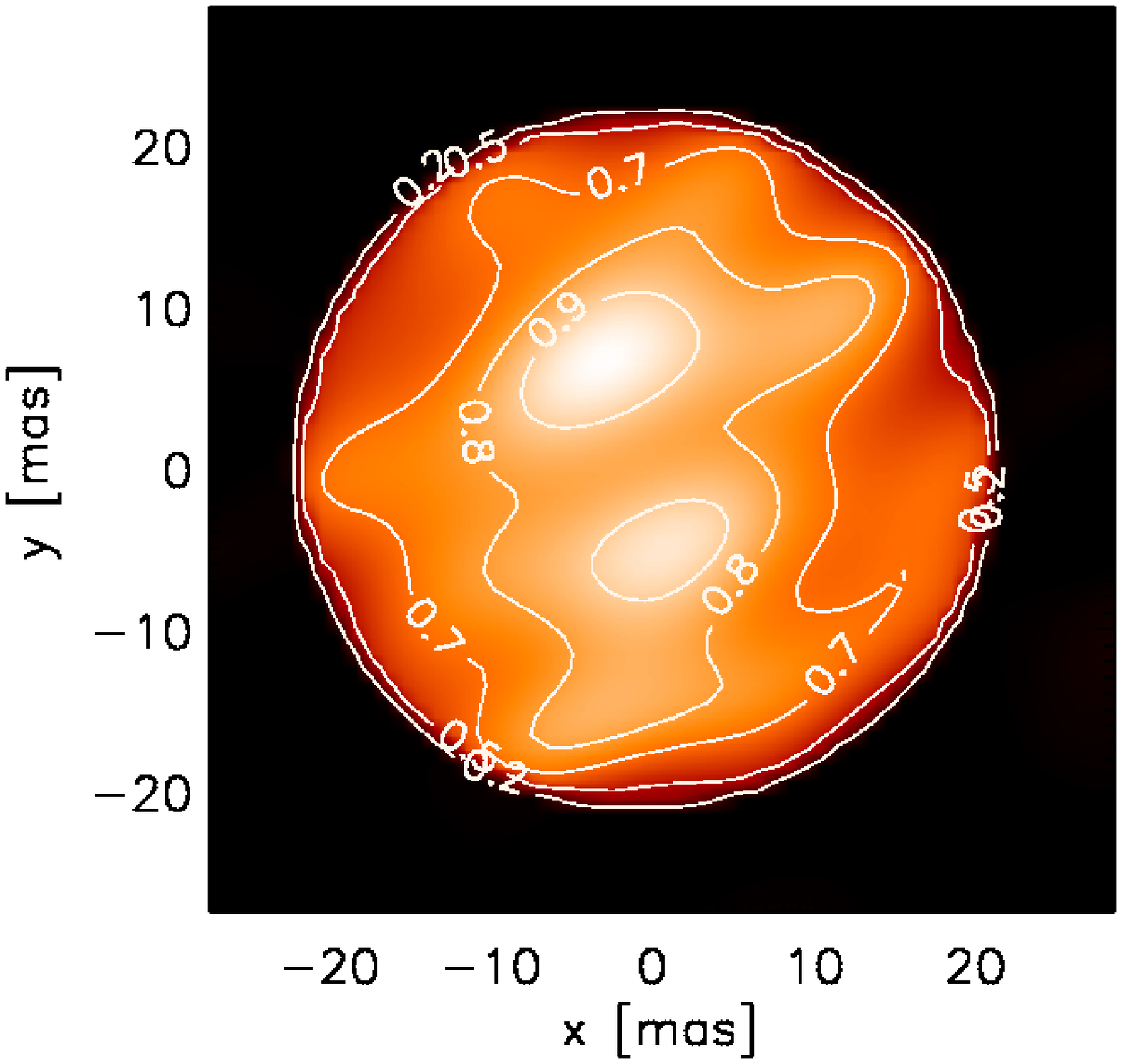}
       \includegraphics[width=0.32\hsize]{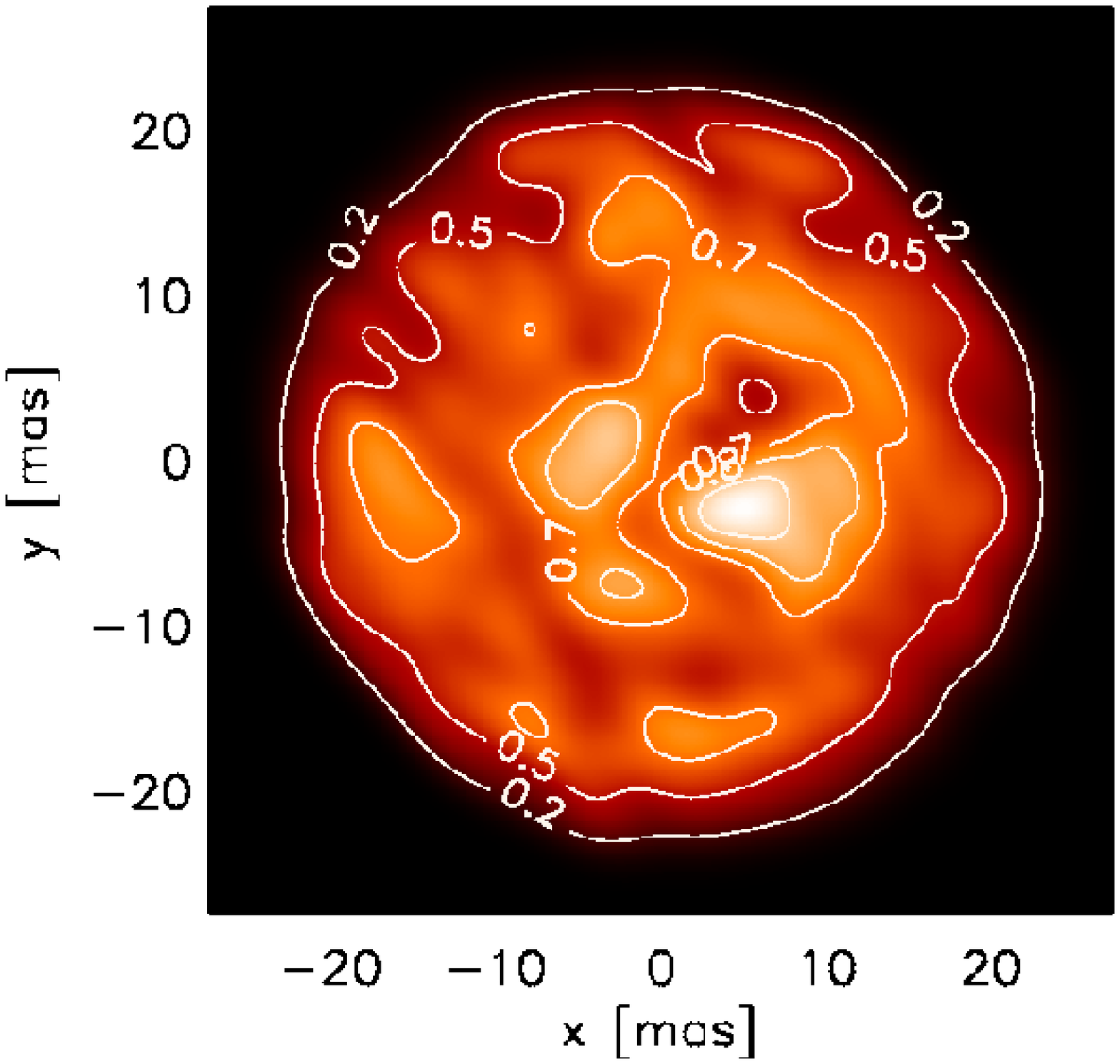}
 \end{tabular}
\caption{\emph{Left panel:} recontructed image from
  \cite{2009A&A...508..923H}. \emph{Right panel:} our best fitting 3D
  simulation snapshot of Fig.~\ref{comparison_1b} convolved with a
  6$\times$6 mas PSF derived by fitting the central peak of the
  interferometric dirty beam.  The intensities in both panels are
  normalized to the range [0, 1] and some contour lines are indicated
  (0.2, 0.5, 0.7, 0.8, 0.9 of the peak brightness).}
              \label{images_1}%
\end{figure*}

This is the first robust confirmation of the physical origin of
surface granulation for Betelgeuse, following on from the detection in
the K band (Paper~I).  Recently, \cite{2009A&A...508..923H} were able
to reconstruct two images of Betelgeuse, using the data presented in
this work, with two different image reconstruction algorithms. The
image reconstructed with WISARD \citep{2009M, 2005JOSAA..22.2348M} is
displayed in Fig.~\ref{images_1} (left). Both reconstructed images
in \citeauthor{2009A&A...508..923H} paper
have two spots of unequal brightness located at roughly the same
positions near the center of the stellar disk. One of these spots is
half the stellar radius in size. Fig.~\ref{images_1} shows a
comparison of the reconstructed image to our best fitting snapshot of
Fig.~\ref{comparison_1b}. Fainter structures are visible in the
synthetic image (right panel) while the reconstructed image (left
panel) is dominated by two bright spots. Moreover, the bigger spot
visible in the reconstructed image is not present in our synthetic
image, whereas there is good agreement in term of location with the smaller spot
located close to the center. However, it is possible that the synthetic map does not match
exactly the location of the spots because it cannot perfectly reproduce
the closure phase data. 

\subsubsection{Molecular contribution to the visibility curves}

\begin{figure*}
   \centering
  \begin{tabular}{ccc}
       \includegraphics[width=0.32\hsize]{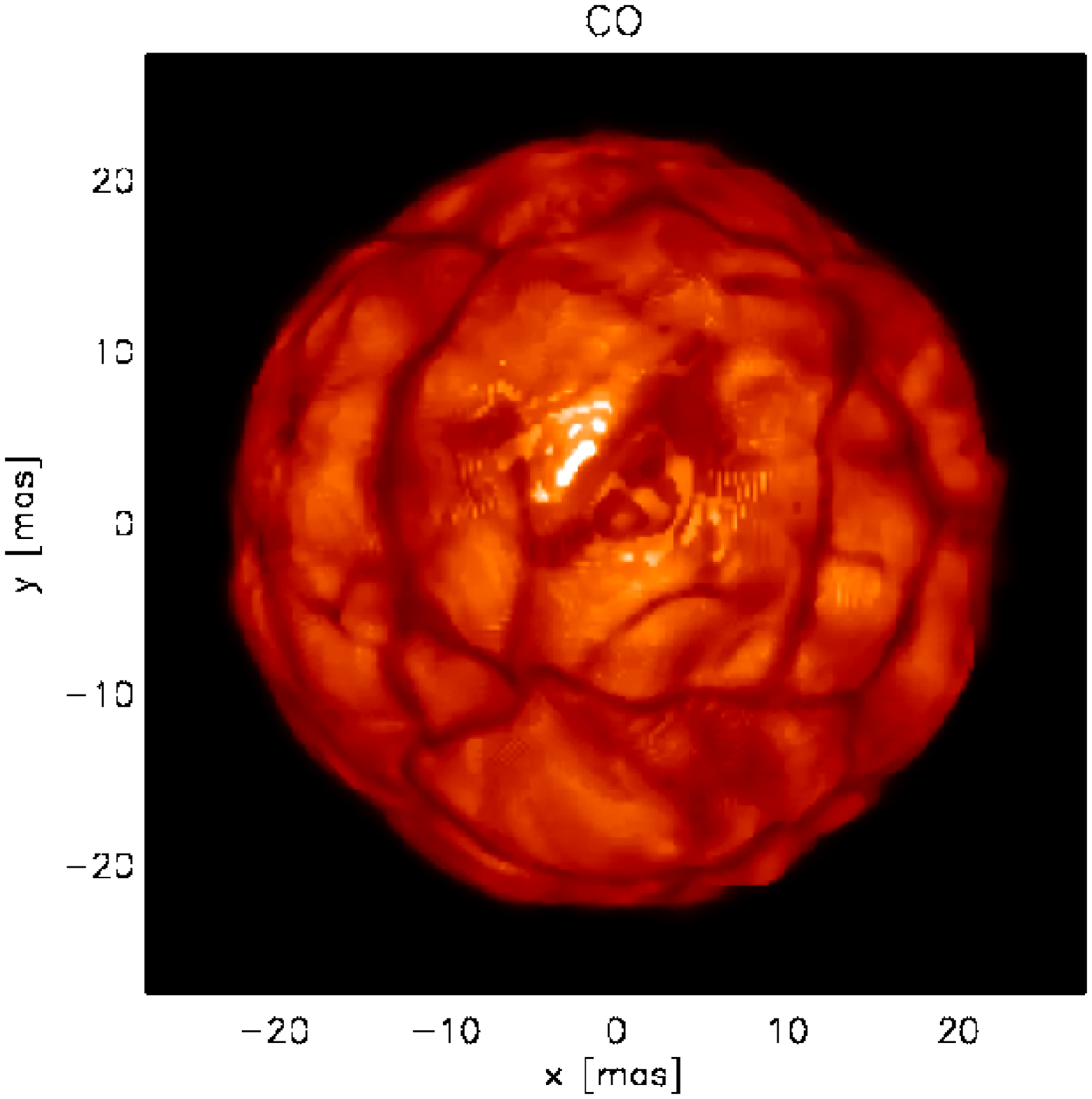}
       \includegraphics[width=0.32\hsize]{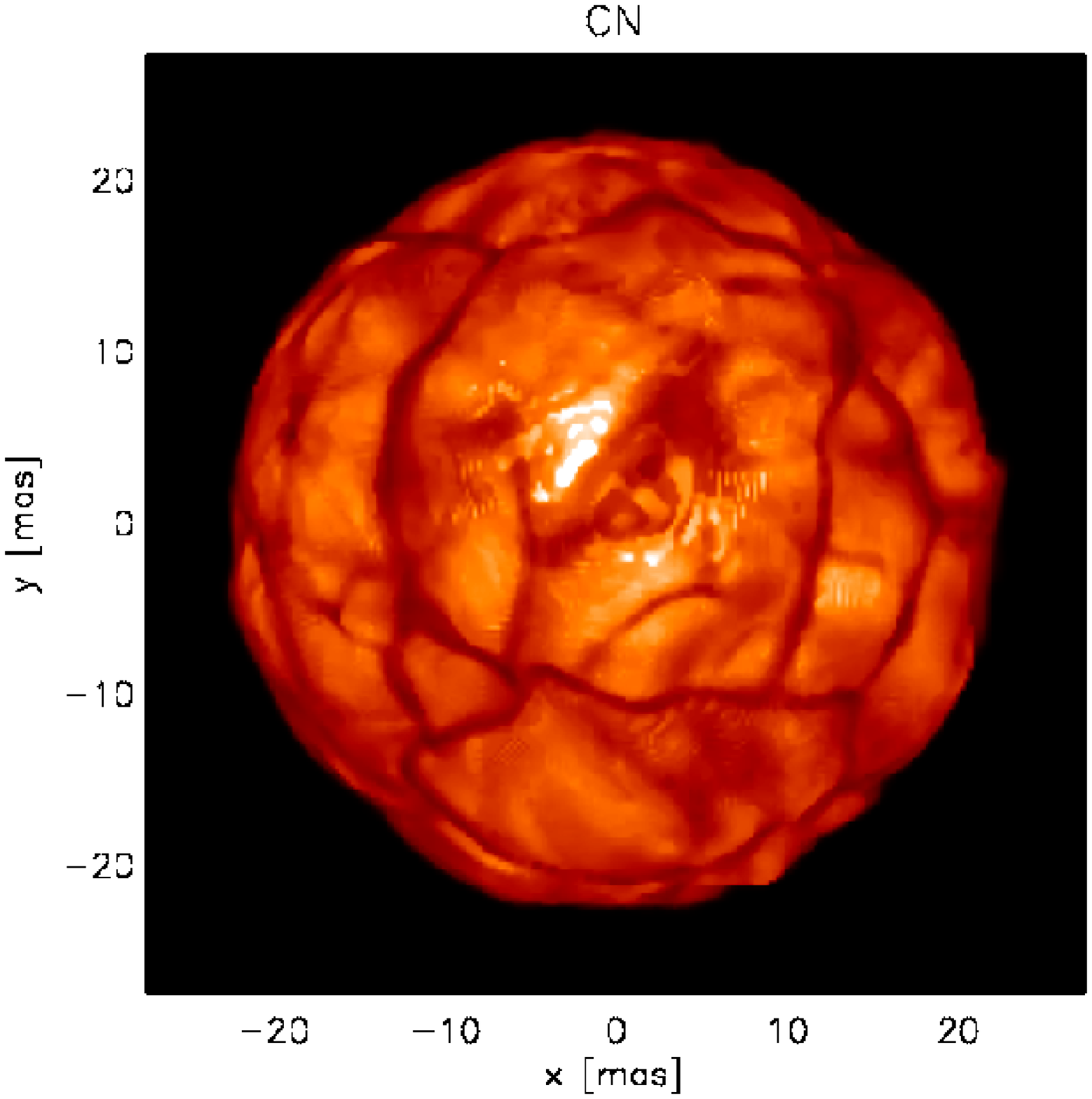}
       \includegraphics[width=0.32\hsize]{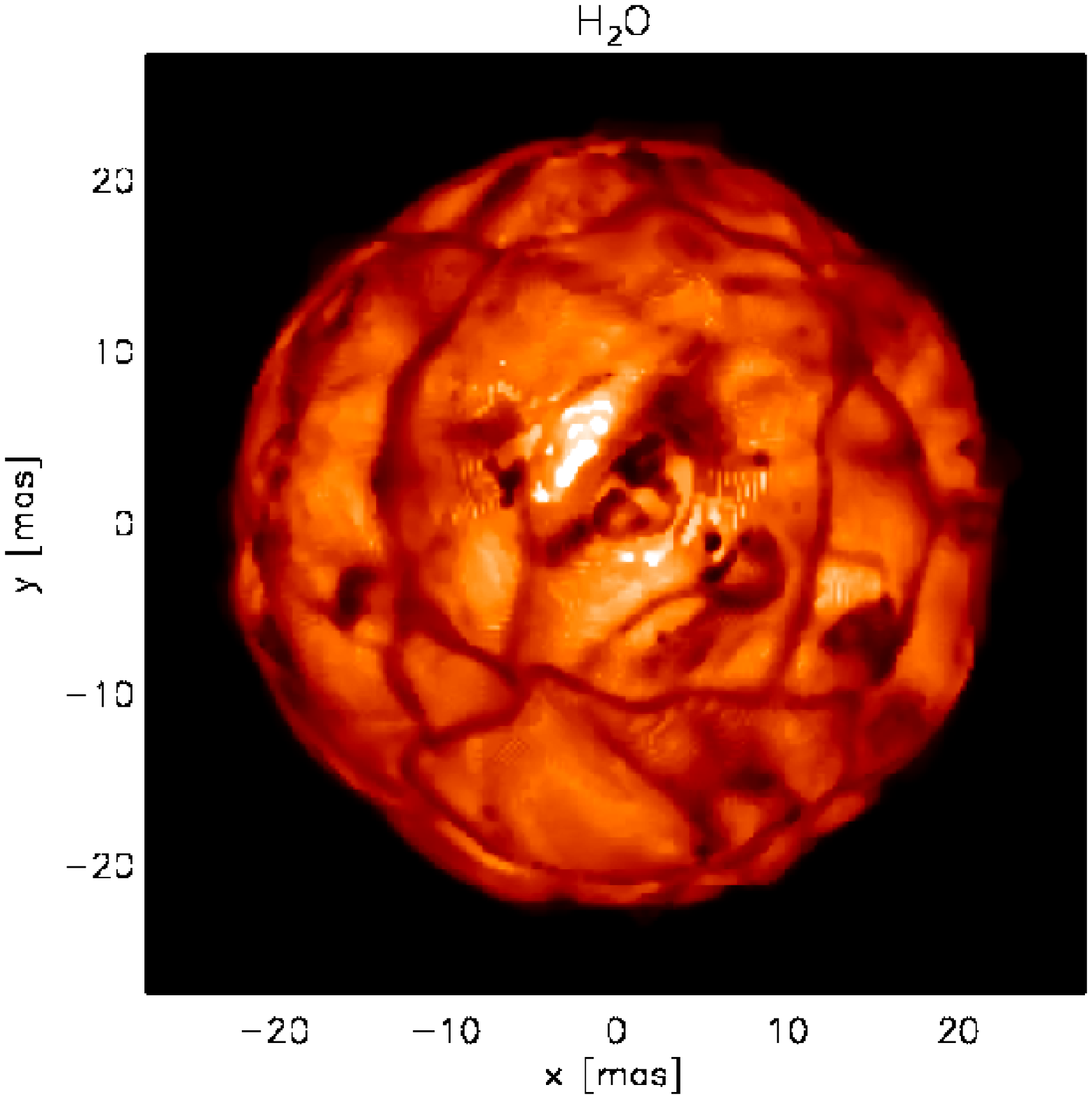}\\
\includegraphics[width=0.5\hsize]{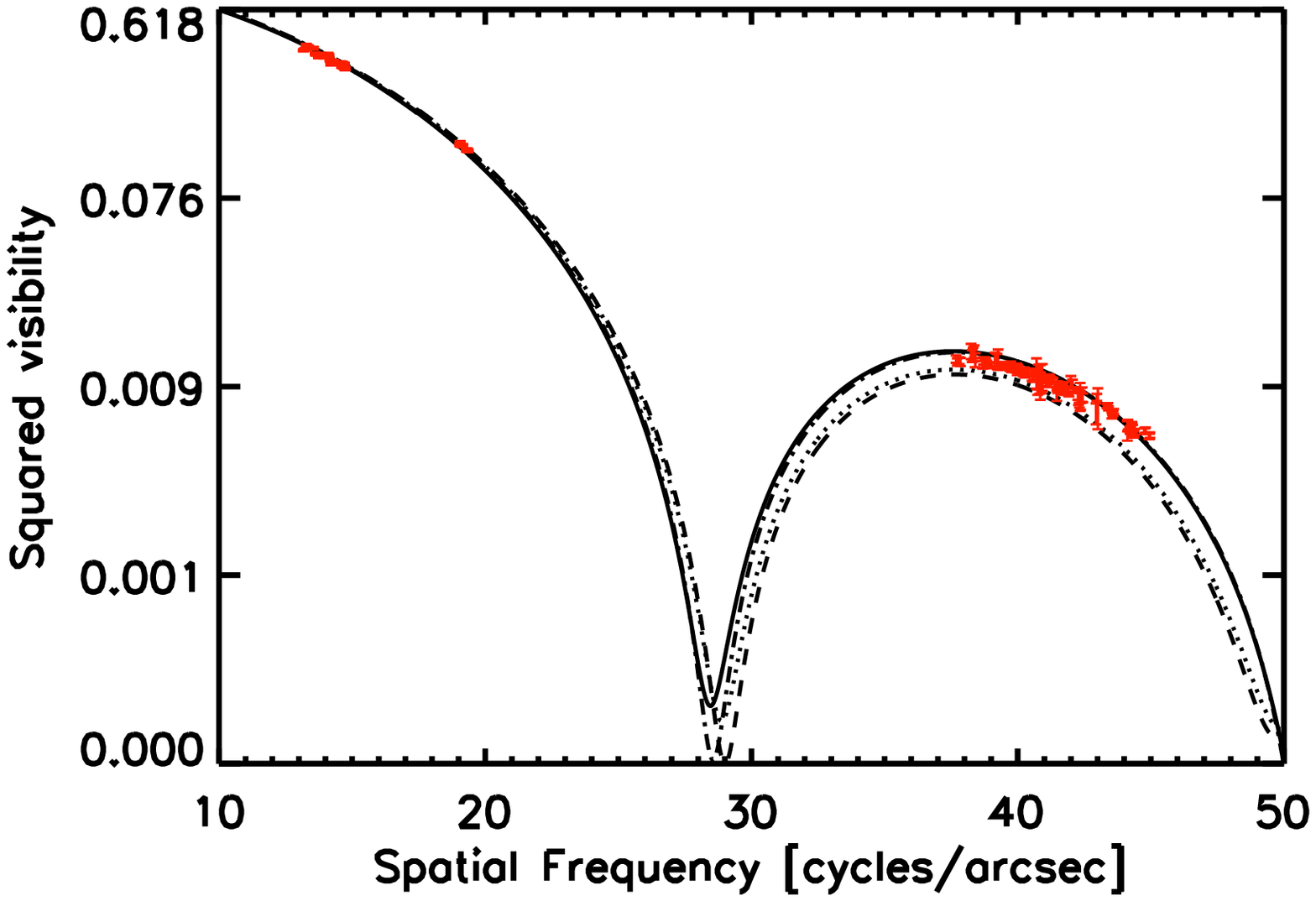}
     \includegraphics[width=0.5\hsize]{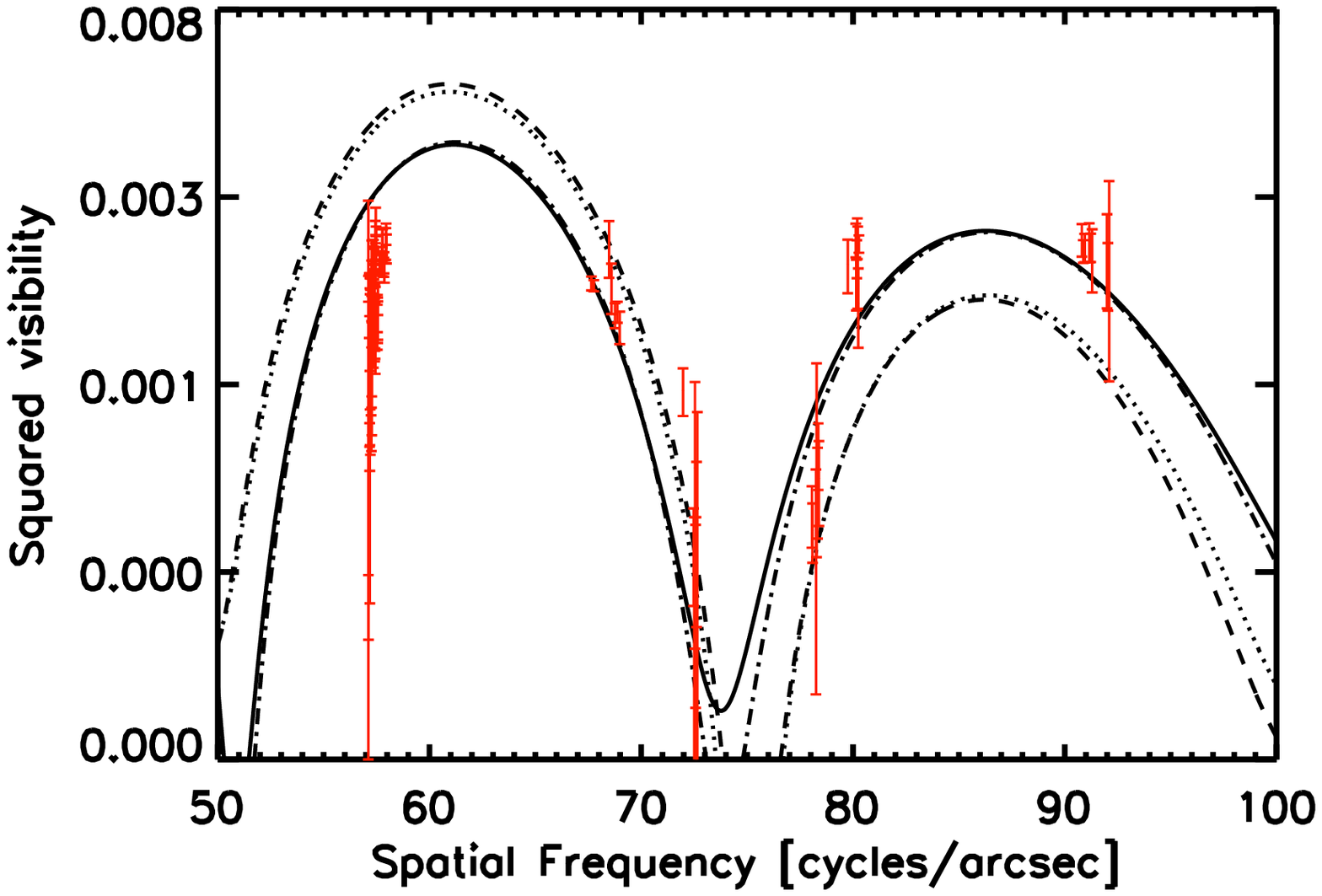}
 \end{tabular}
\caption{\emph{Top row:} maps of the linear intensity in the IONIC
  filter of the molecular species (Fig.~\ref{filters}). The
  simulation snapshot and intensity range is the same as in
  Fig.~\ref{comparison_1b}. \emph{Bottom row:} the best
  matching visibility curve (solid line) of Fig.~\ref{comparison_1a} is
  compared to the visibility curve obtained from the intensity maps
  above: CO (dotted line), CN (dashed line), and  H$_2$O (dash dot
  line). Red data points are IOTA observations.
}
              \label{molecules}%
\end{figure*}

It is important to determine what molecular species contribute the
most to the intensity absorption in the stellar atmosphere. For this
purpose, we used the best fitting snapshot of Fig.~\ref{comparison_1b}
and recomputed the intensity maps in the IONIC filter using only
CO, CN and H$_2$O molecules, because they are the largest absorbers at
these wavelengths (Fig.~1 of this work and Fig.~3 of Paper~I). The intensity maps
displayed in Fig.~\ref{molecules} of this paper (top row) should be
compared to the original one in Fig.~\ref{comparison_1b} which accounts
for all the molecular and atomic lines. The surfaces of CO and CN maps
clearly show the granulation pattern and they are spot-free. However,
the H$_2$O map shows dark spots which can be also identified in the
original intensity map.

We also calculated visibility curves from these molecular intensity
maps using the same rotation angle used to generate the synthetic data
in Fig.~\ref{comparison_1a}. Figure~\ref{molecules} (bottom row) shows
that the H$_2$O visibility is the closest to the original one both at
low and high spatial frequencies. We conclude that: (i) in the first
lobe, the H$_2$O visibility is smaller than the CO and CN
visibilities. Thus the radius of the star is dependent on the H$_2$O
contribution. (ii) At higher frequencies, only the H$_2$O visibility
can fit the observed data whereas the CO and CN visibilities fit
poorly.

\subsubsection{Size distribution on the stellar surface}

We also characterized the typical size distribution on the
stellar surface using interferometric observables. The large range of
spatial frequencies of the observation (between 12 to 95
arcsec$^{-1}$) is very well suited for this purpose.

Out aim is to visualize the energy within the signal as a
function of spatial frequency. After the computation of the Fourier
Transform, $FT$, we obtain
$\hat{I}\left(u,v\right)=FT\left[I\left(x,y\right)\right]$ where
$I\left(u,v\right)$ is the best matching intensity map of
Fig.~\ref{comparison_1b}. The resulting complex number
$\hat{I}\left(u,v\right)$ is multiplied by low-pass and high-pass
filters to extract the information from different spatial frequency ranges
(corresponding to the visibility lobes). Finally an inverse Fourier Transform,
$\bar{FT}$, is used to obtain the filtered image:
$I_{\rm{filtered}}\left(x,y\right)=\bar{FT}\left[\hat{I}\left(u,v\right)\cdot\rm{filter}\right]$.

\begin{figure*}
   \centering
  \begin{tabular}{cc}
\includegraphics[width=0.32\hsize]{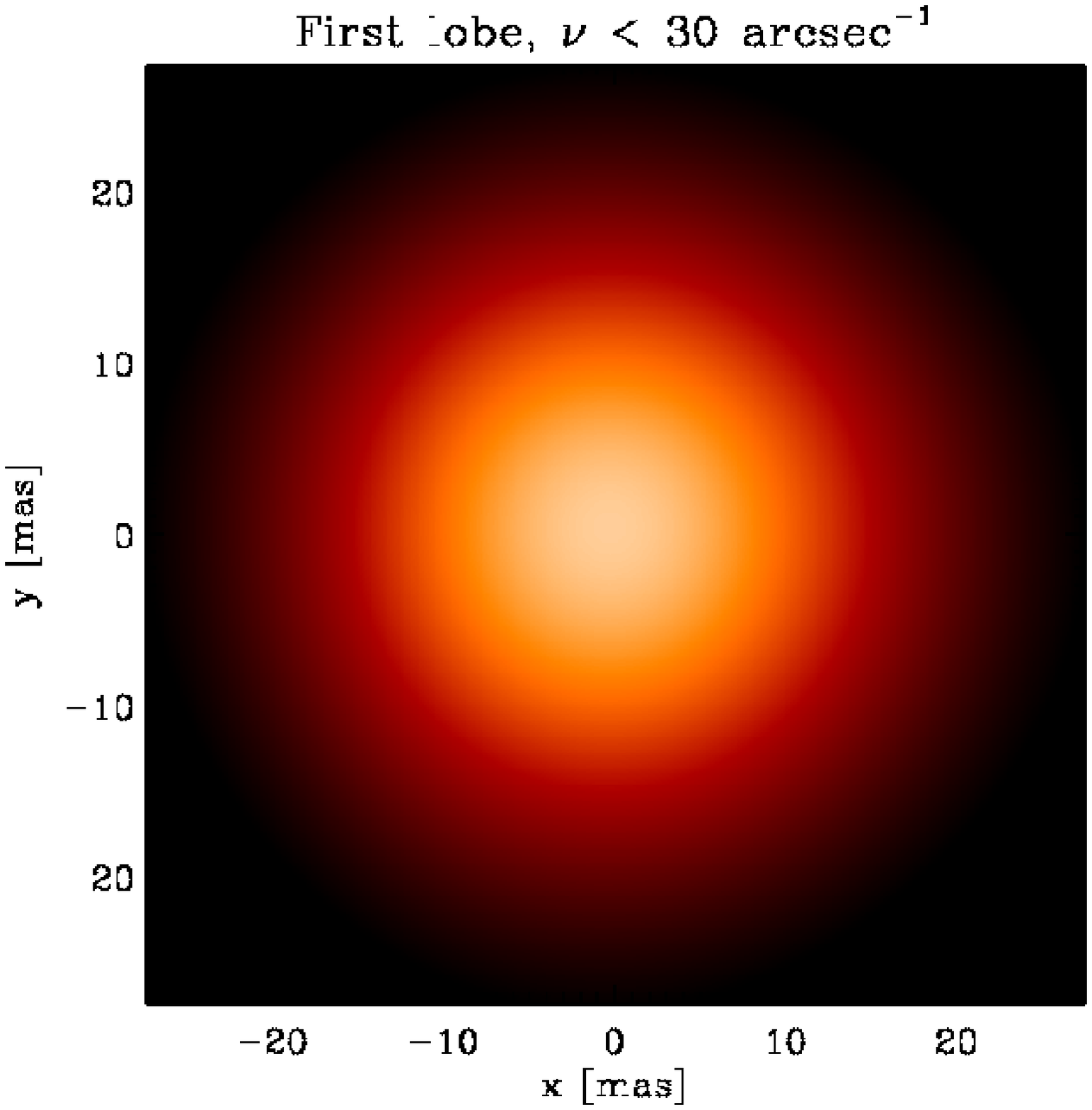}
     \includegraphics[width=0.32\hsize]{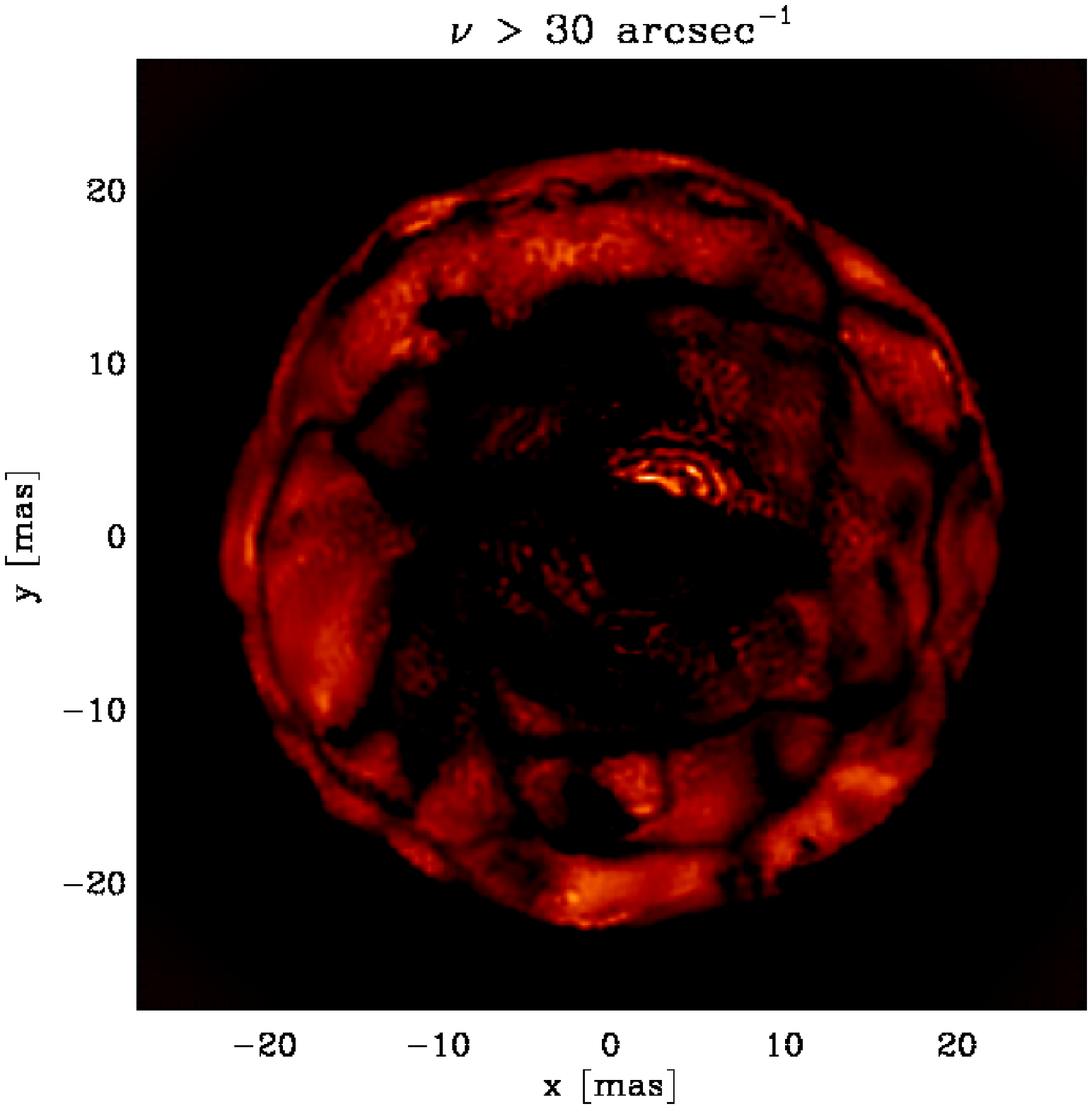}\\
       \includegraphics[width=0.32\hsize]{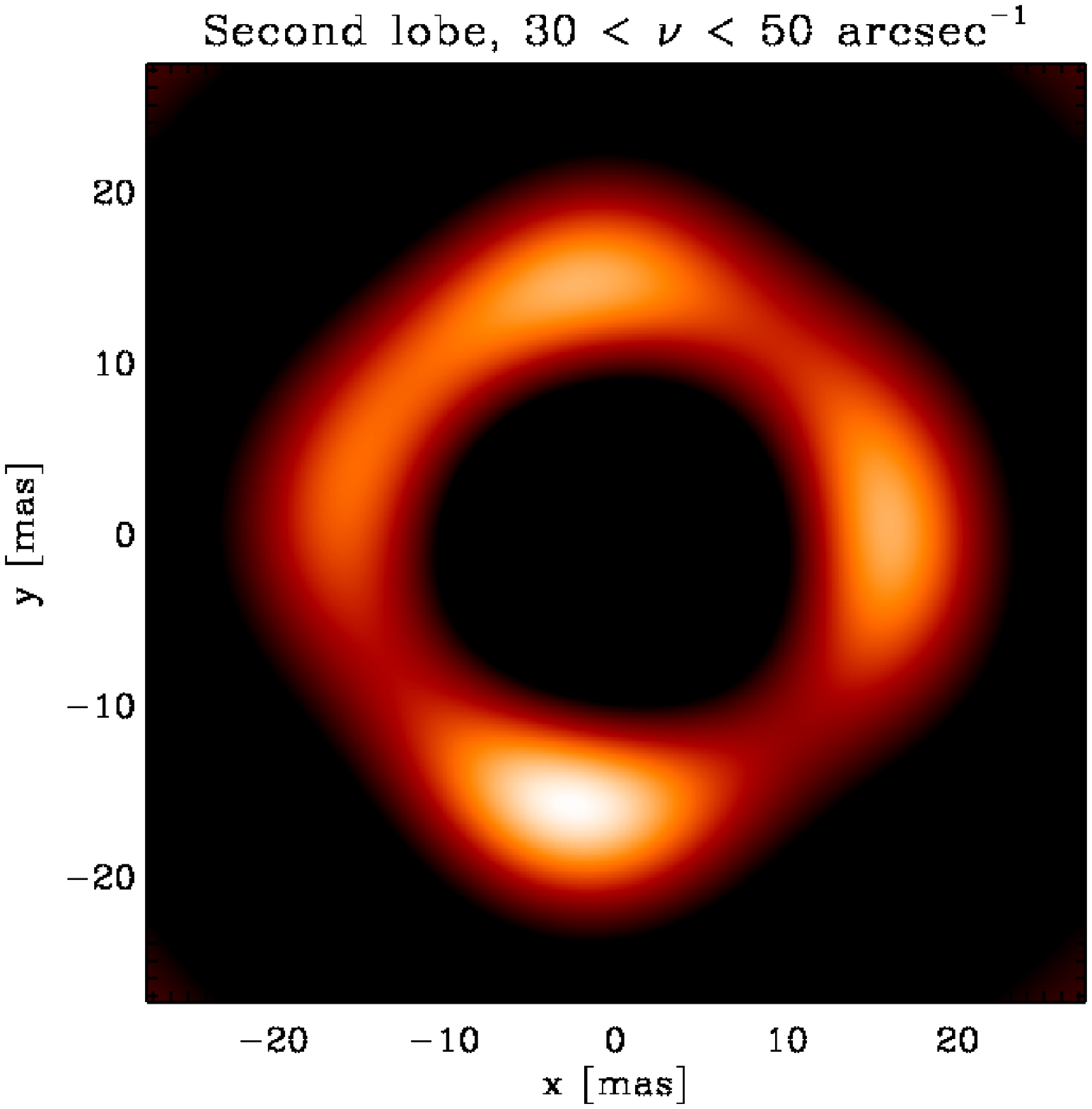}
       \includegraphics[width=0.32\hsize]{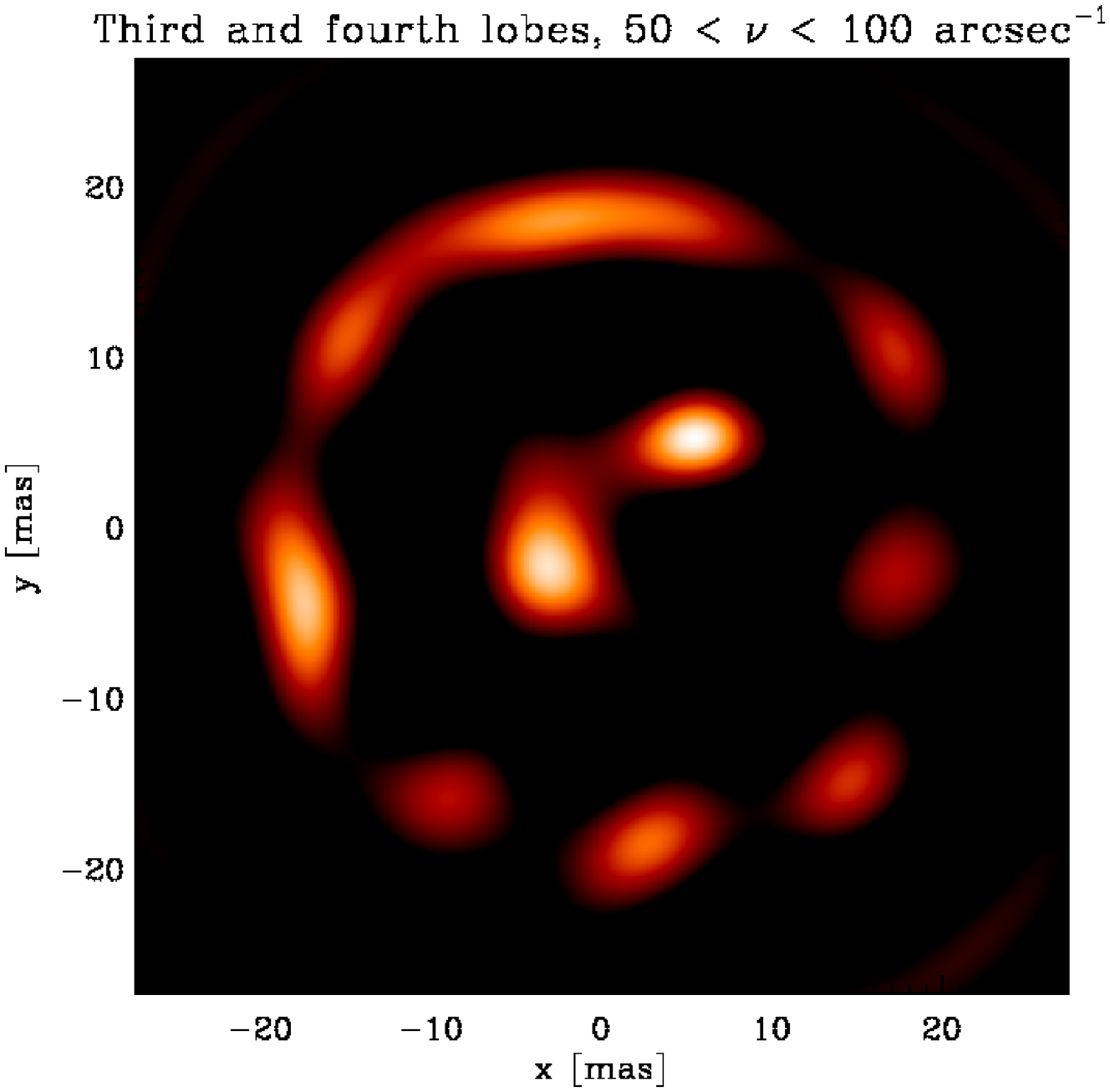}
 \end{tabular}
\caption{Intensity maps filtered at different spatial frequencies
  corresponding to the lobes of the visibility curve shown in
  Fig.~\ref{comparison_1a}. The images are normalized between [0,1].
}
              \label{size}%
\end{figure*}

Fig.~\ref{size} (top left panel) shows the filtered image at spatial
frequencies, $\nu$, corresponding to the first lobe. Due to the fact
that we cut off the signal at high spatial frequencies, the image
appears blurry and seems to contain only the information about the
stellar radius. However, the top right panel displays the signal related
to all the frequencies larger than the first lobe: in this image we
clearly miss the central convective cell of $\approx$ 30 mas size ($60\%$
of the stellar radius) visible in Fig.~\ref{comparison_1b}. Thus, the
first lobe also carries information on the presence of large convective
cells.

Fig.~\ref{size} (bottom row) shows the second lobe with
convection-related structures of $\approx$ 10-15 mas, ($30\%$ of the
stellar radius), and the third and fourth lobes with structures
smaller than 10 mas.  We conclude that we can estimate the presence of
convection-related structures of different size using visibility
measurements at the appropriate spatial frequencies. However, only
imaging can definitively characterize the size of granules. A first
step in this direction has been carried out in Berger et al. 2010 (to
be sumbitted soon), where the image reconstruction algorythms have
been tested using intensity maps from this RHD simulation. In the case
of Betelgeuse, we have fitted its interferometric observables between 12
and 95 arcsec$^{-1}$ and thus inferred the presence of small to medium
scale granules (5 to 15 mas) and a large convective cell ($\approx$ 30
mas).

\subsection{Data from 7000 to 12900 \AA}

The simulated surface in the optical to near-infrared region shows a
spectacular pattern characterized by dark spots and bright areas. The
brightest areas can be up to 50 times more intense than the dark
ones. In addition, this pattern changes strongly with time and has a
lifetime of a few weeks. In the wavelength region below
$\approx 1$ $\mu$m, the resulting surface pattern, though related to
the granulation below, is also connected to dynamical effects. In fact, the light
comes from higher up in the atmosphere where the optical depth is
smaller than one and where waves and shocks start to dominate. In
addition to this, the emerging intensity depends on the opacity run
through the atmosphere and veiling by TiO molecules is very strong at
these wavelengths (Fig.~\ref{filters}).

The filters centered in the optical part of this wavelength range are
characterized by the presence of strong molecular lines while the
infrared filter probes layers closer to the continuum-forming
region. Observations at wavelengths in a molecular band and in the
continuum probe different atmospheric depths, and thus layers at
different temperatures. They provide important information on the
wavelength-dependence of the limb-darkening and strong tests of our
simulations.

Since the observations have been made at two different epochs, we
fitted them individually: (i) the data taken in 1997
\citep{2000MNRAS.315..635Y}, and (ii) the data from 2004
\citep{2004young}. We proceeded as in Section \ref{ionicsect}. Note
that the filter curve for the 7820 \AA\ bandpass was lost and so a
top-hat was assumed instead; we tested the validity of this assumption
by replacing the known optical-region filter curves with top-hats,
which did not affect the synthetic visibility and closure phase data
significantly. Within the large number of computed visibilities and
closure phases for each filter, we found that there are two snapshots
of the simulated star, one for each epoch, that fit the
observations. At each epoch the same rotation angle of the snapshot
was found to fit all of the observed wavebands.

\begin{figure*}
   \centering
    \begin{tabular}{ccc}
\includegraphics[width=0.36\hsize]{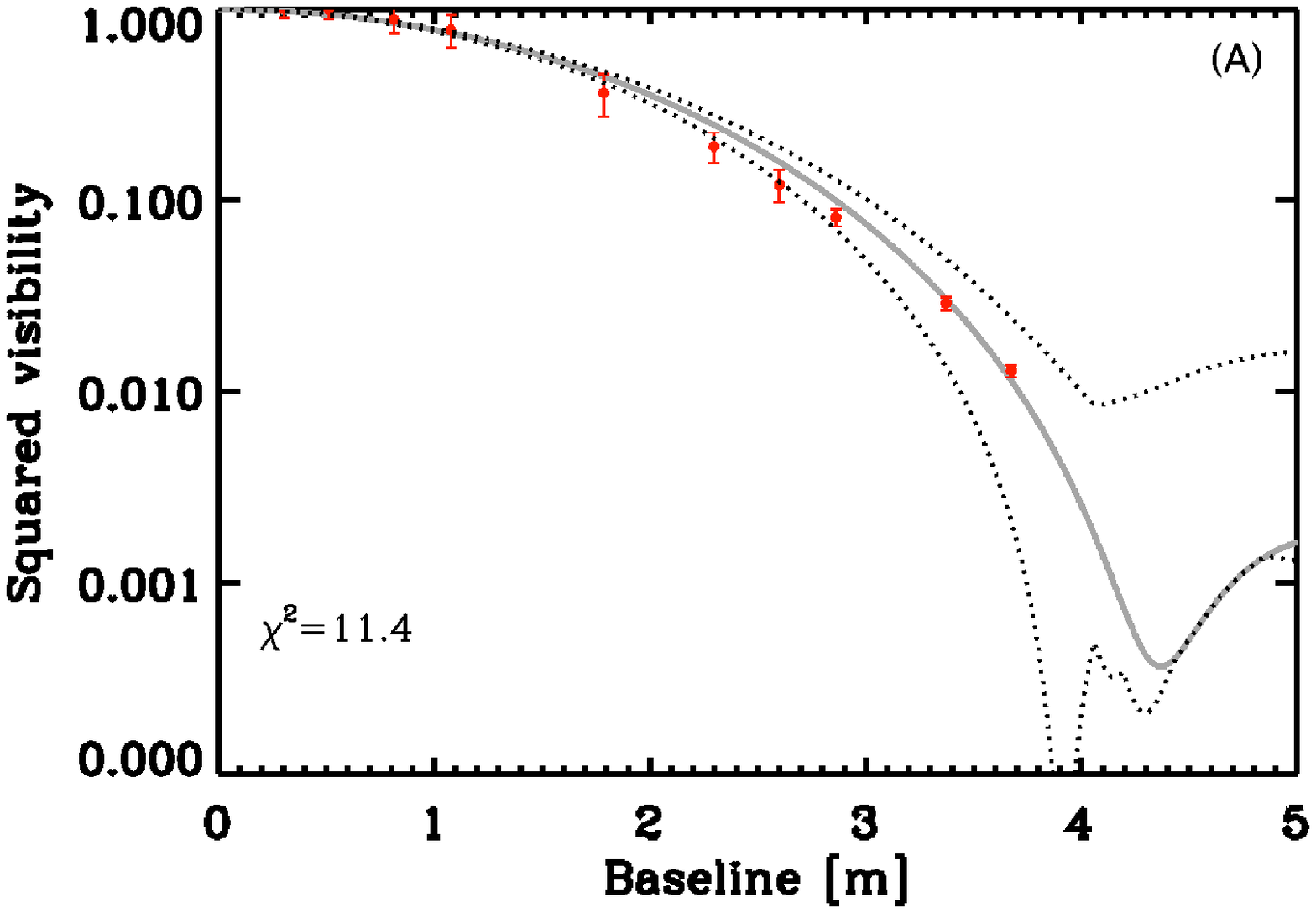}%COAST4
\includegraphics[width=0.36\hsize]{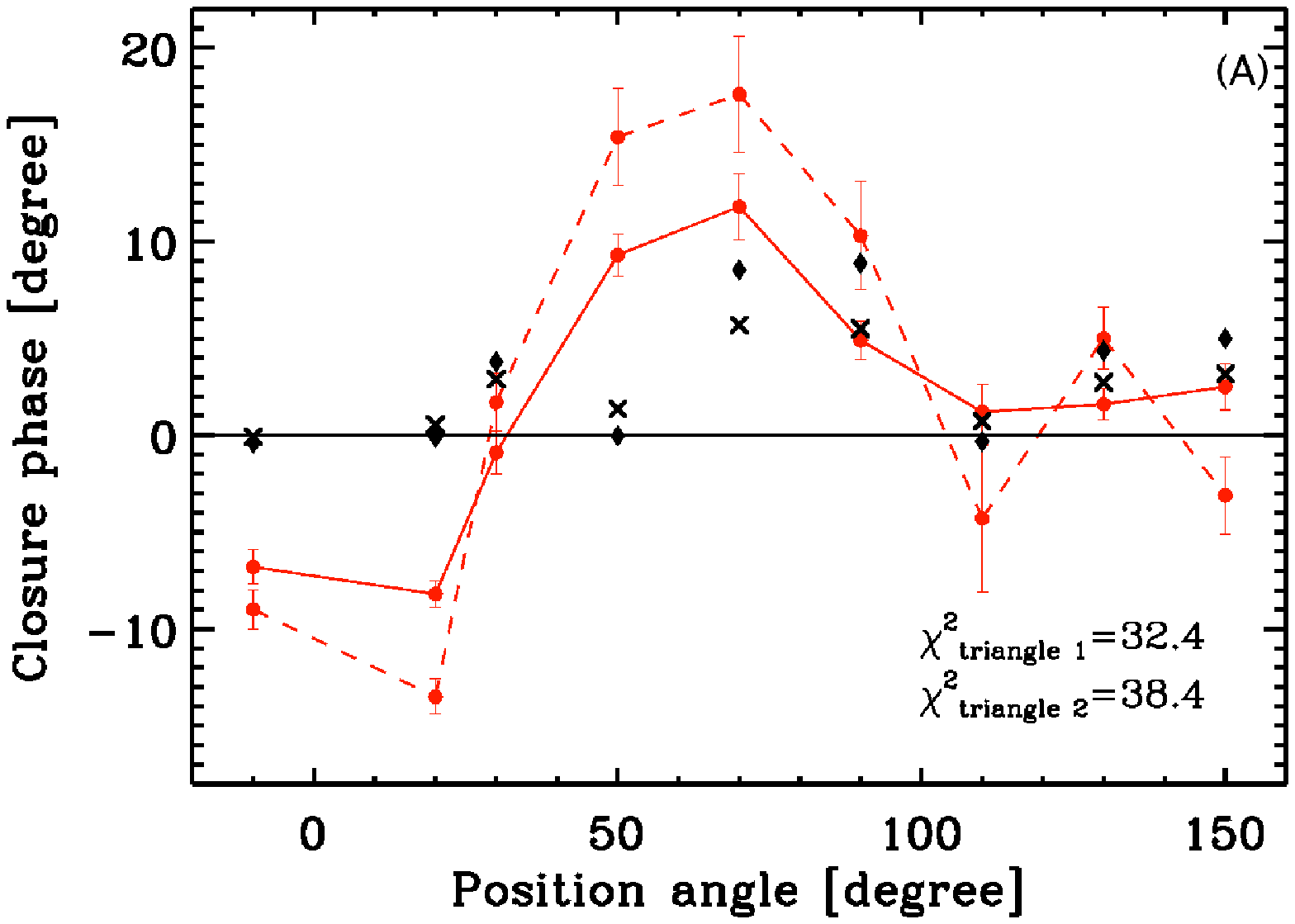}
\includegraphics[width=0.27\hsize]{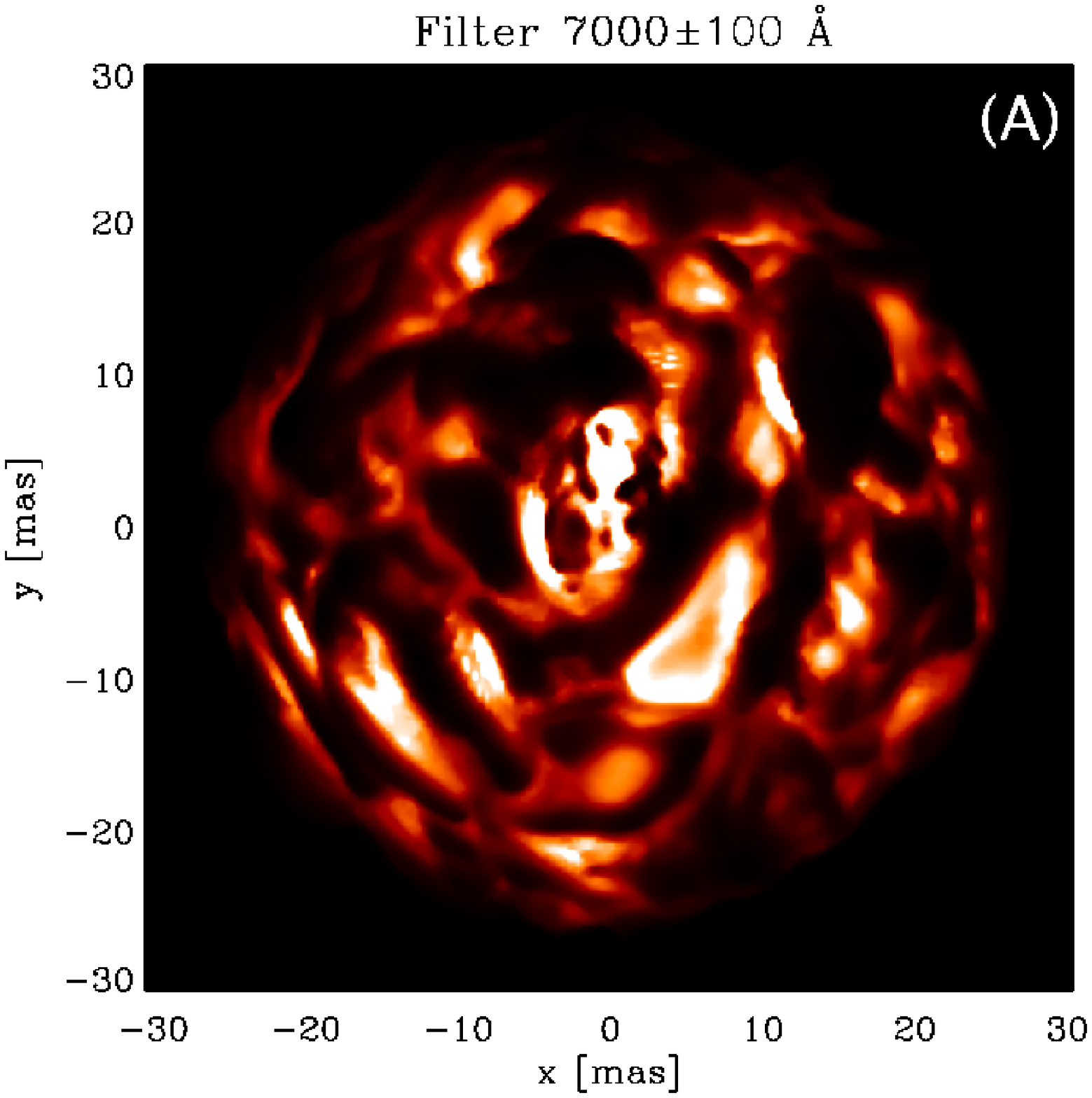}\\
\includegraphics[width=0.36\hsize]{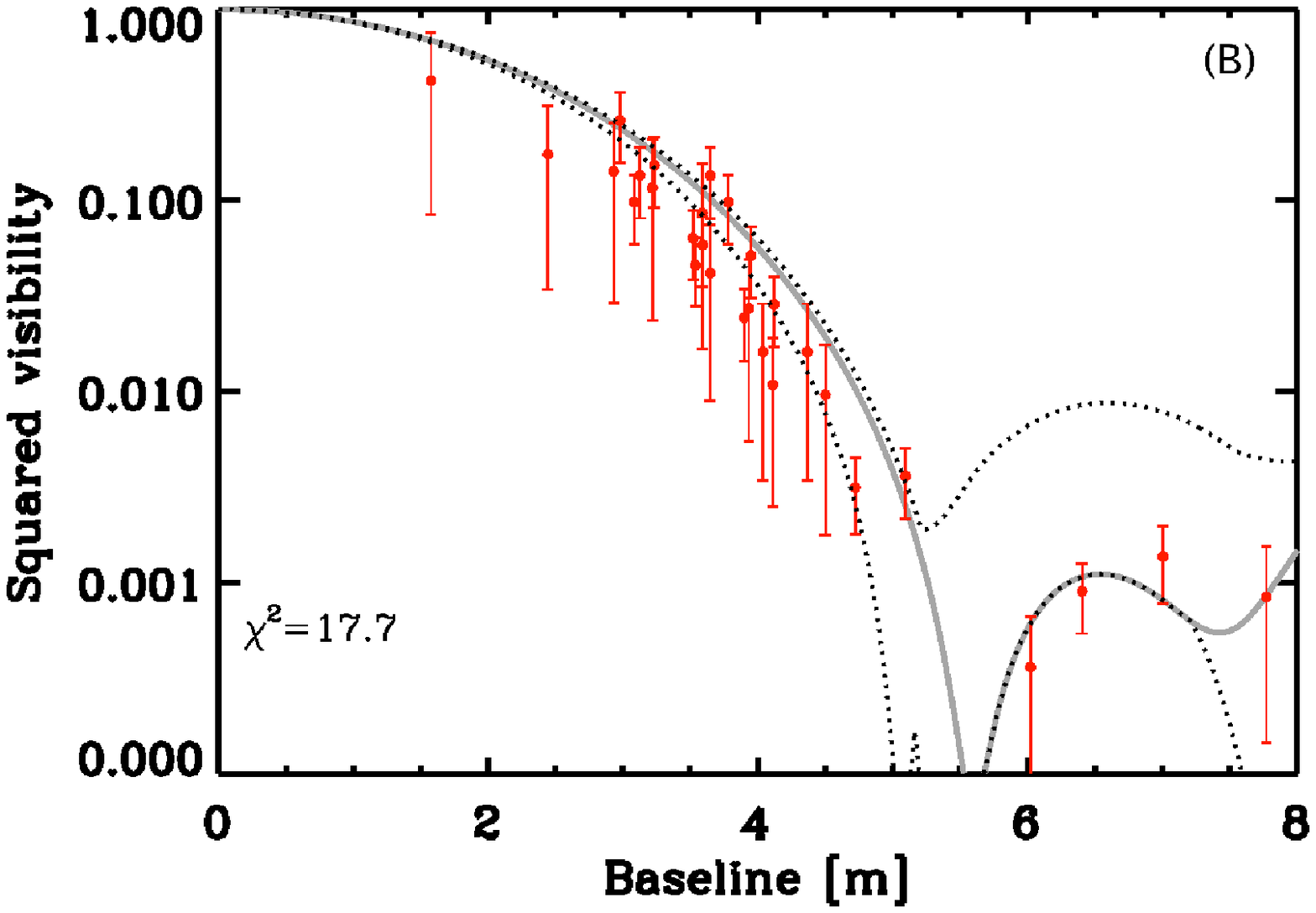}%COAST2
	\includegraphics[width=0.36\hsize]{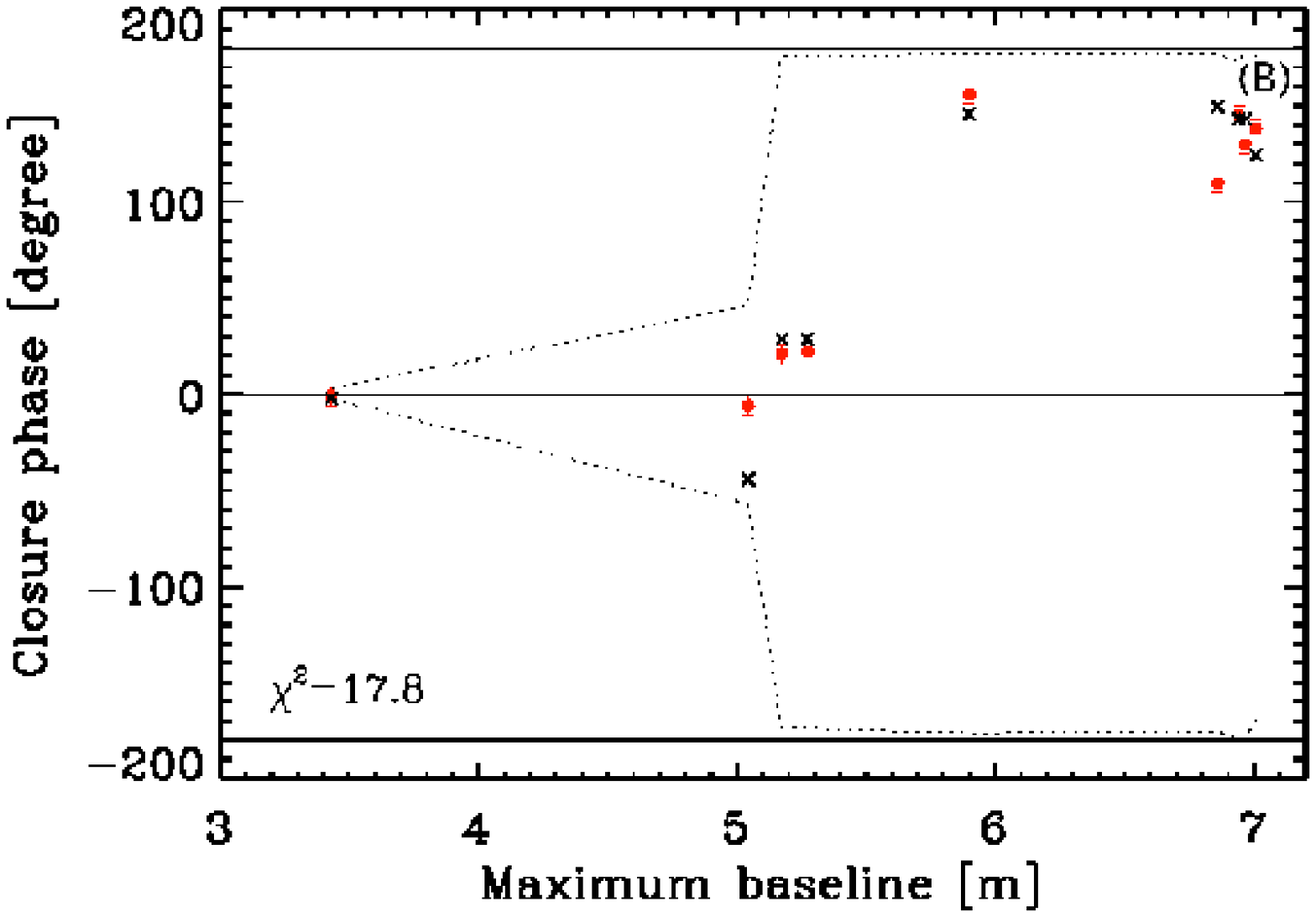}
        \includegraphics[width=0.27\hsize]{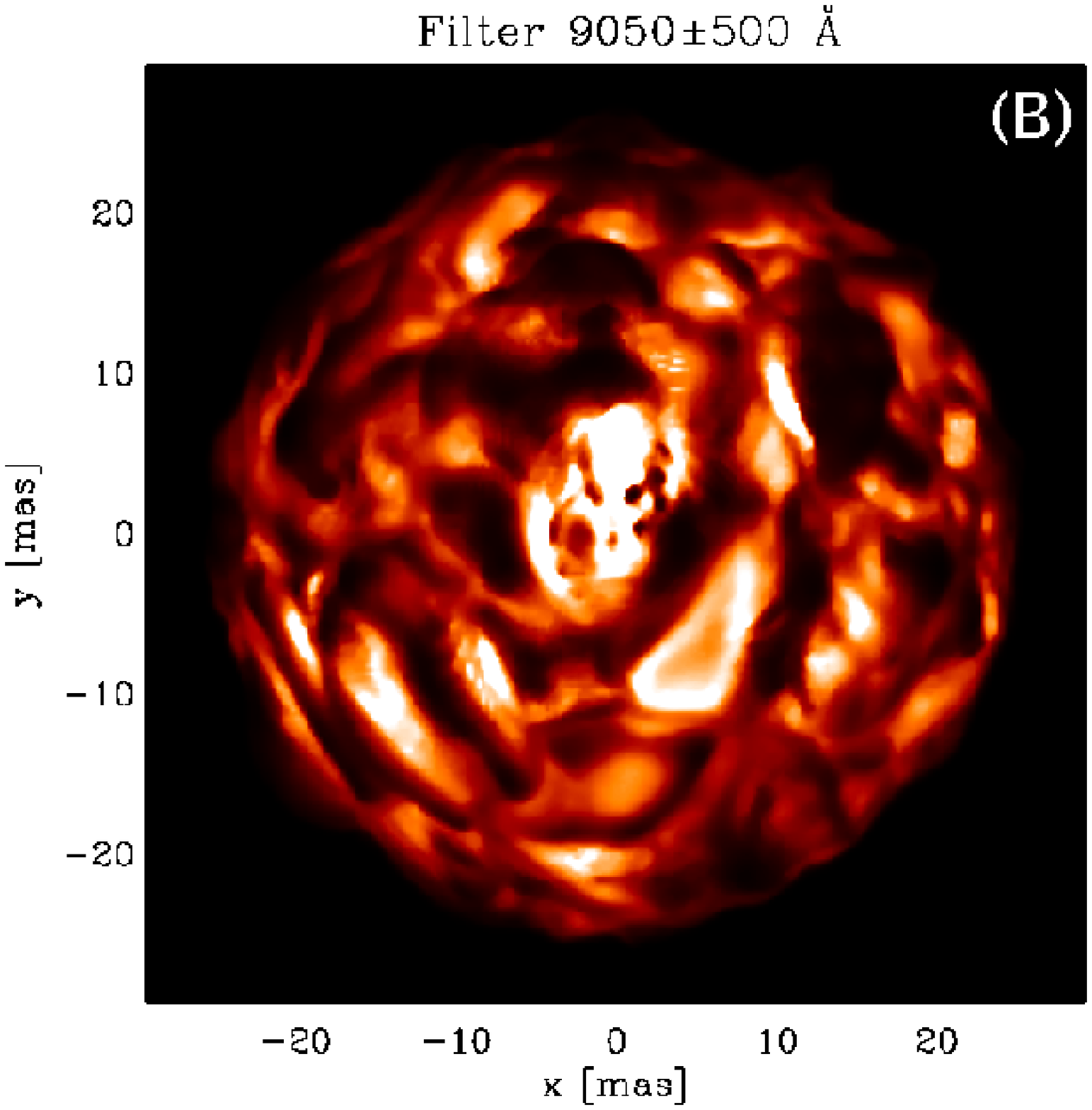}\\
        \includegraphics[width=0.36\hsize]{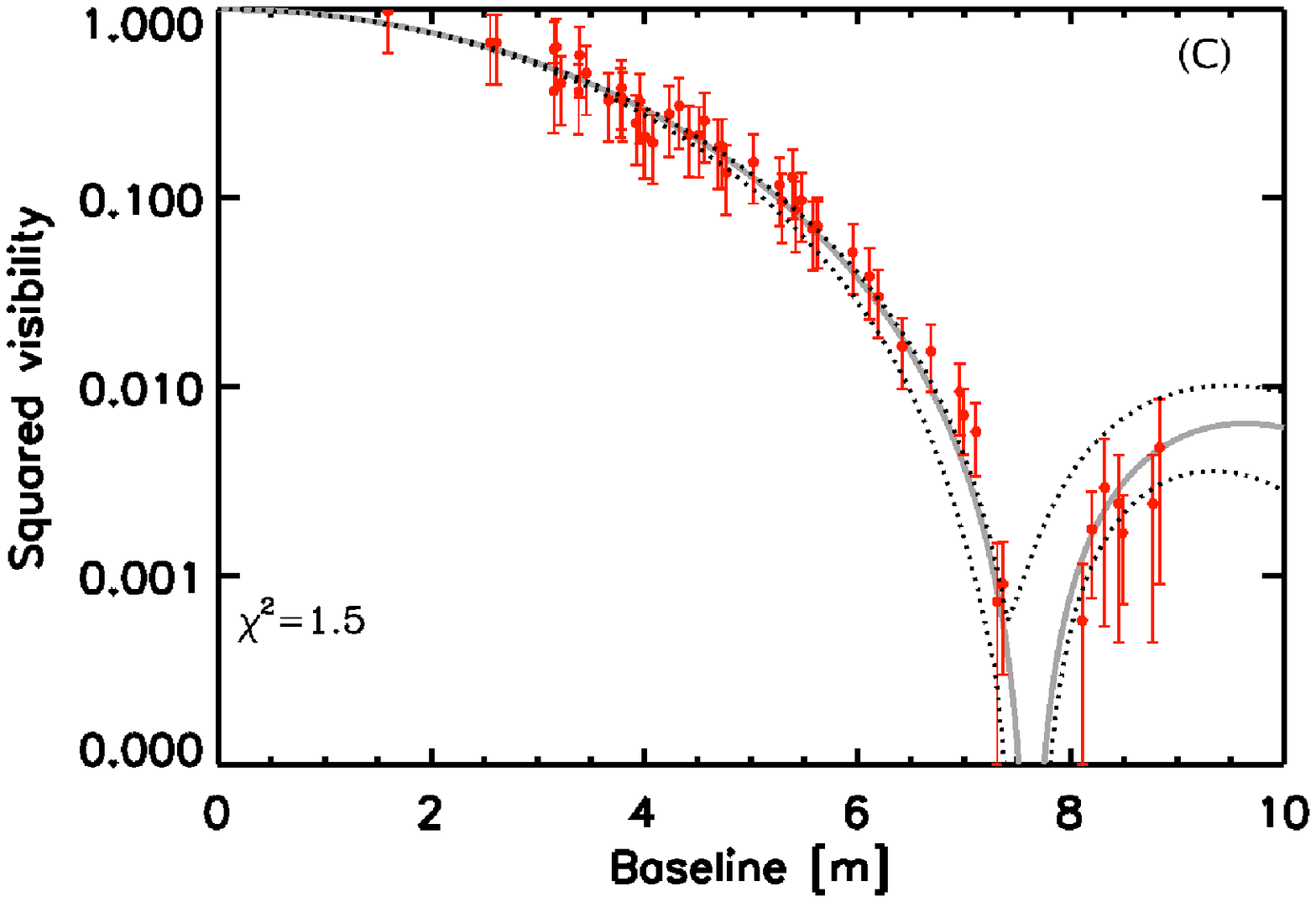}%COAST1
        \includegraphics[width=0.36\hsize]{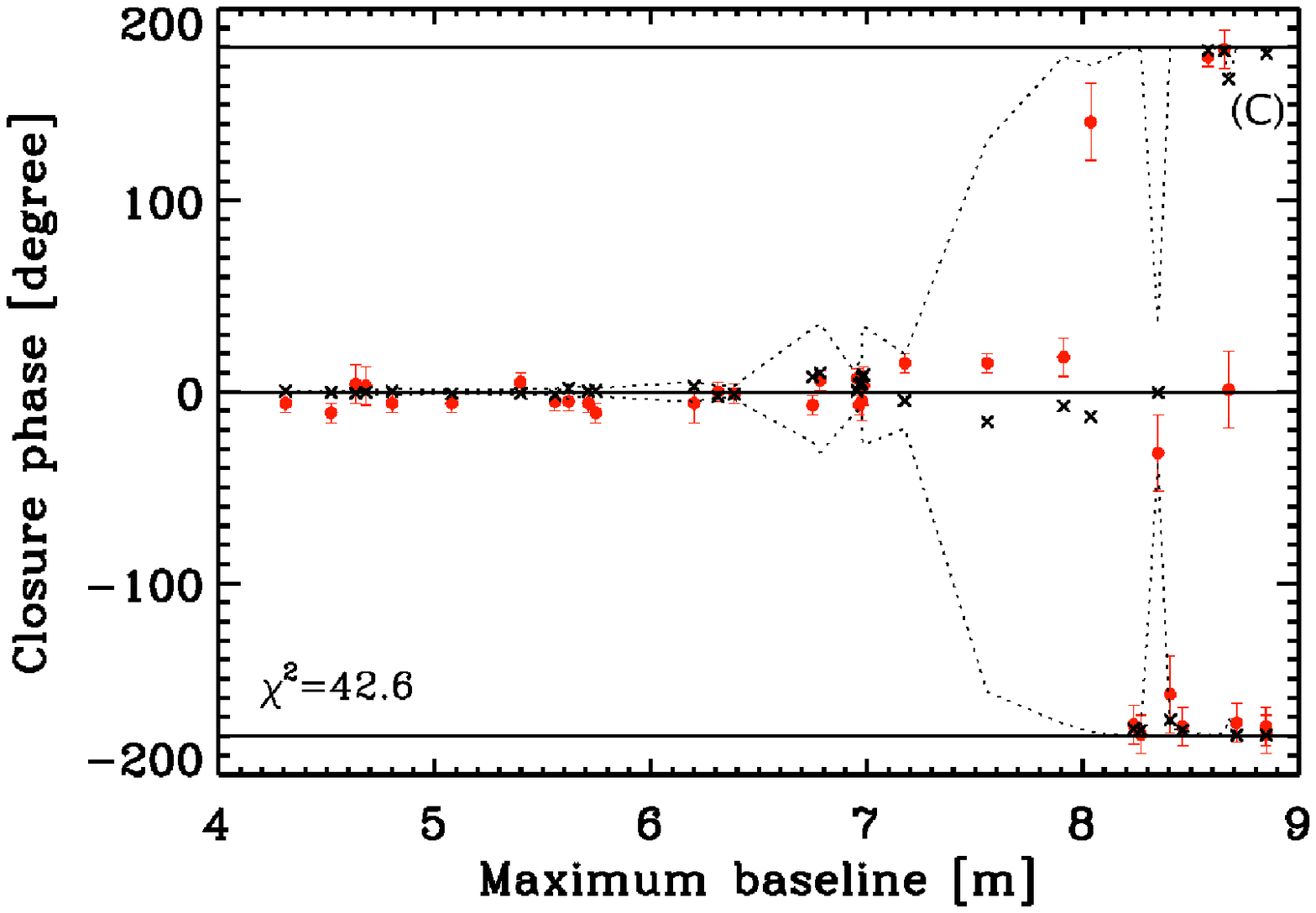}
        \includegraphics[width=0.27\hsize]{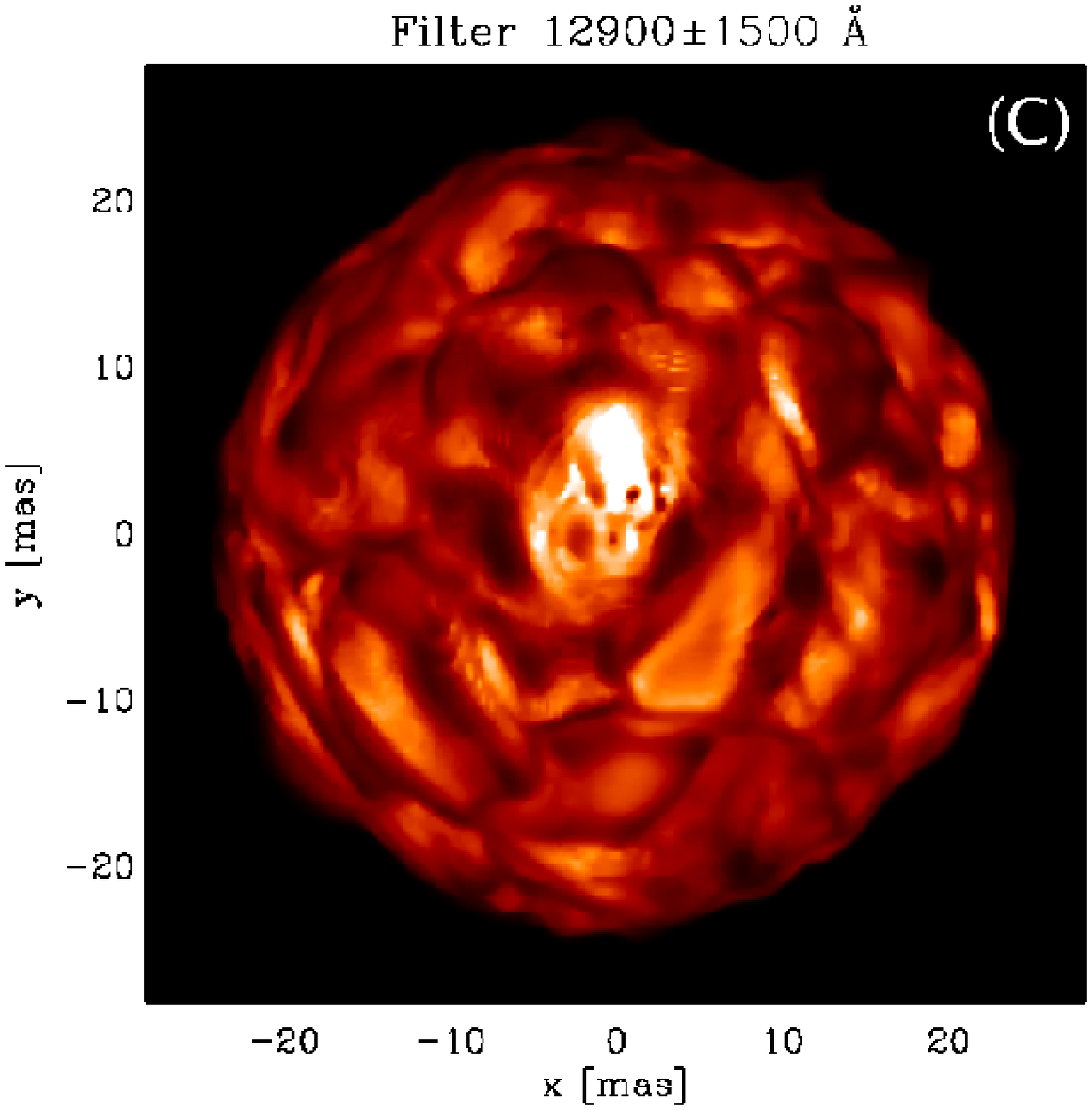}    
\end{tabular}
        \caption{Comparison to the data taken in 1997
          \citep{2000MNRAS.315..635Y}. The symbols in the visibility
          curves and closure phases plots have the same meaning as in
          Fig.~\ref{comparison_1a}. The intensity are normalized
          between [0;
            $3.5\times10^5$]\,erg\,cm$^{-2}$\,s$^{-1}$\,{\AA}$^{-1}$. The
          stellar parameters of this snapshot are: $L =
          93\,500\,L_\odot$, $R = 843.3\,R_\odot$, $T_{\rm eff} =
          3475$\,K and $\log(g) = -0.35$. The row marked with (A),
            (B) and (C) correspond to bandpass centered at 7000, 9050
            and 12900 \AA , respectively. The visibility curves are
            plotted in left column, while the closure phases are shown
          in central column. The intensity maps centered at $7000$ and
          $9050$ \AA \ have been scaled to an apparent diameter of
          $\sim$48.9 mas, at a distance of 161.1 pc fitting the data
          points in the first lobe; the map at $12900$ \AA \ to an
          apparent diameter of $\sim$47.1 mas, at a distance of 167.1
          pc. \emph{Row marked with (A):} the observed visibility amplitudes (90
          visibilities) have been averaged over
the position angle of the
aperture mask for the sake of clarity. We performed the fitting
procedure for each of the unaveraged visibilities, and then we plot
the averaged value. The closure phases of two (triangle 1
            in solid and triangle 2 in dashed red line) of the 10 observed
            triangles have been plotted as a function of the position
            angle of the aperture mask on the sky. The best-matching
            synthetic closure phases are black crosses for triangle 1
            and black diamonds for triangle 2.}
         \label{young_comparison1}
   \end{figure*}

Fig.~\ref{young_comparison1} displays the comparison to the data taken
in 1997. The 7000 \AA\ synthetic image corresponds to a region with
strong TiO absorption (transition ${\rm A}^3\Phi - {\rm  X}^3\Delta (\gamma)$, see Fig.~\ref{filters}). This is also true for the
9050 \AA\ image (transition ${\rm E}^3\Pi - {\rm  X}^3\Delta (\epsilon)$) but
in this case the TiO band is less strong. The relative intensity
  of TiO bands depend on the temperature gradient of the model and
  change smoothly from one snapshot to another. The map at 12900 \AA\ is TiO
free and there are mostly CN lines: in this case, the surface
intensity contrast is less strong than in the TiO bands.

\begin{figure*}
   \centering
    \begin{tabular}{ccc}
\includegraphics[width=0.36\hsize]{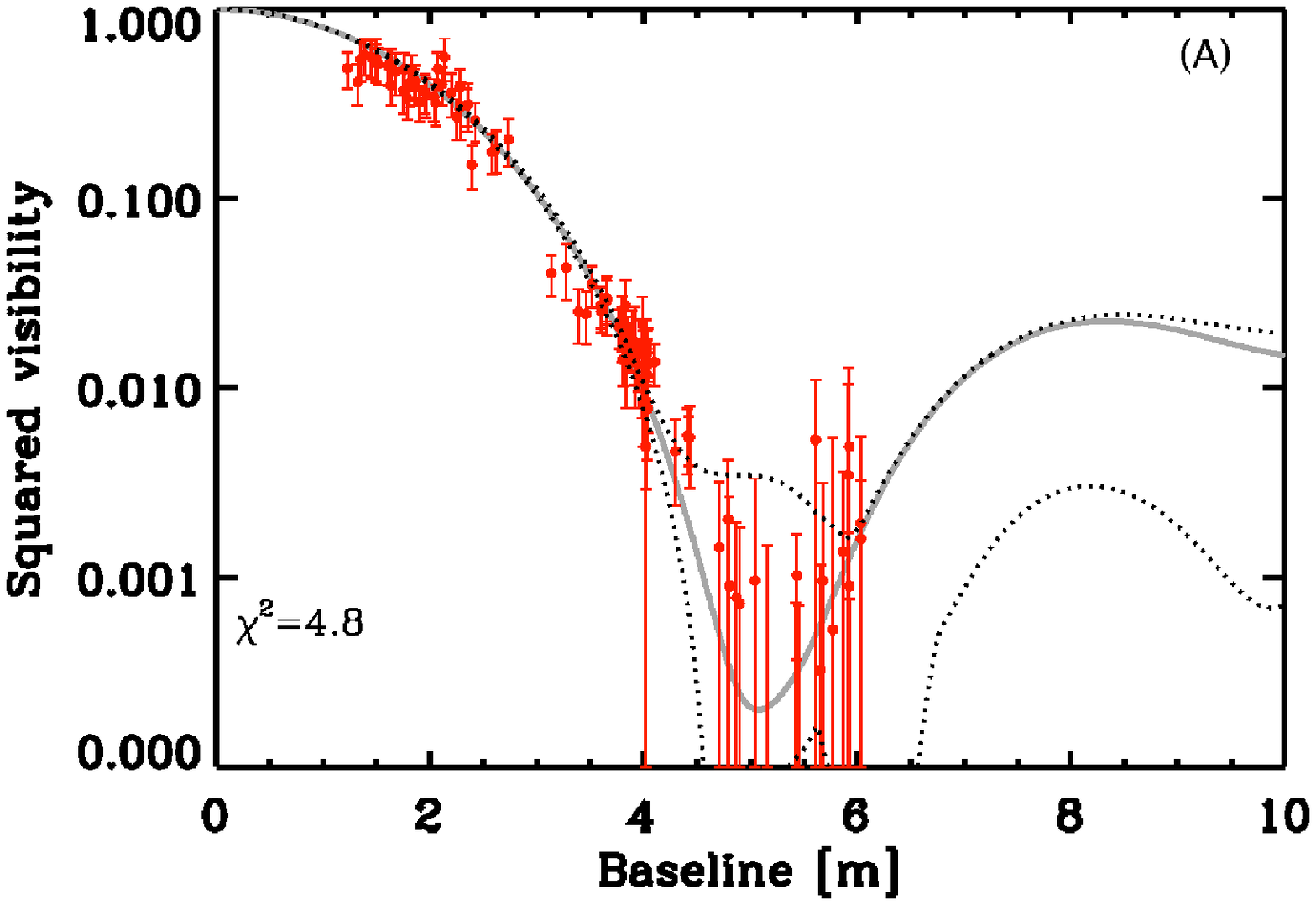}%COAST3
\includegraphics[width=0.36\hsize]{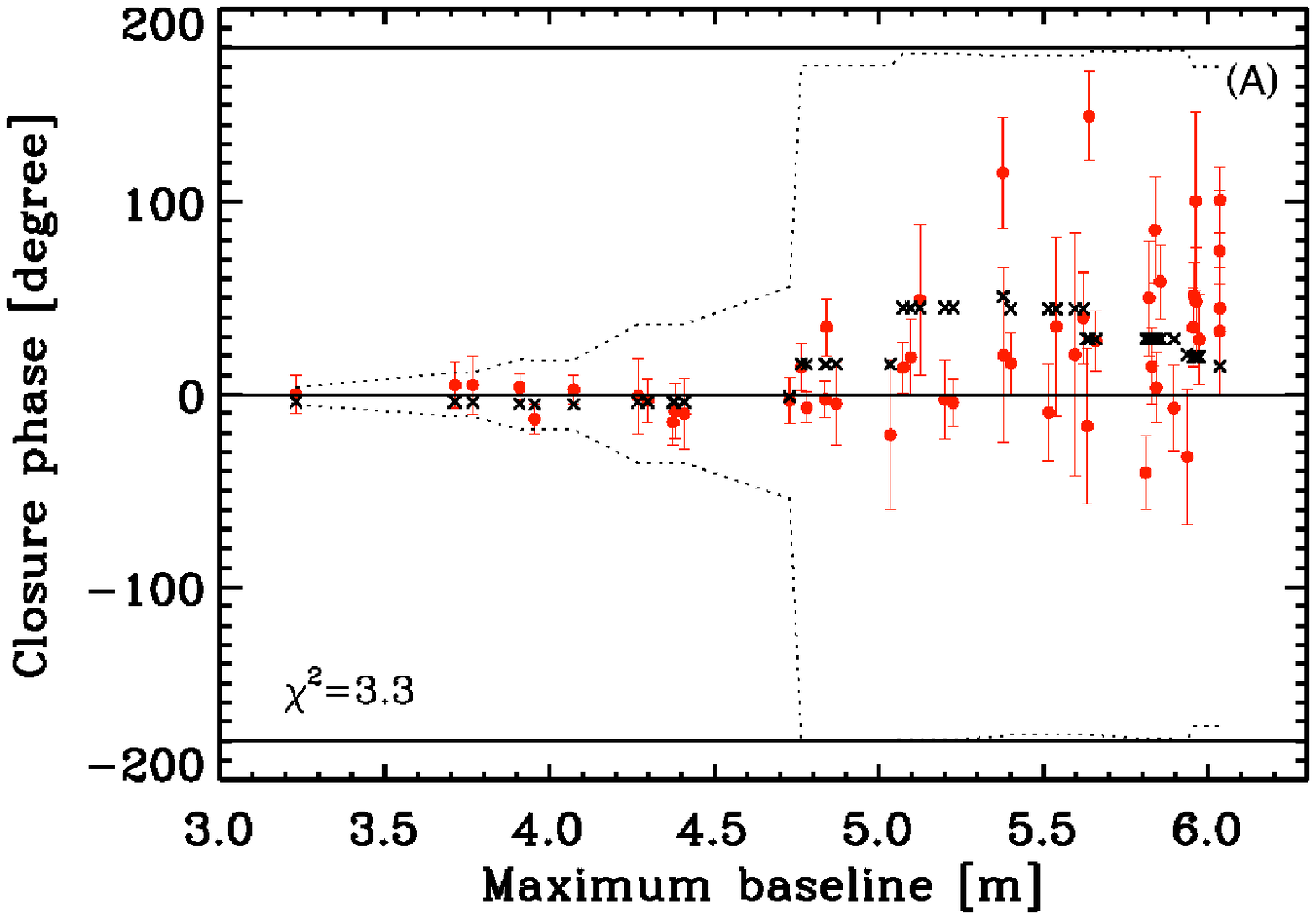}
\includegraphics[width=0.27\hsize]{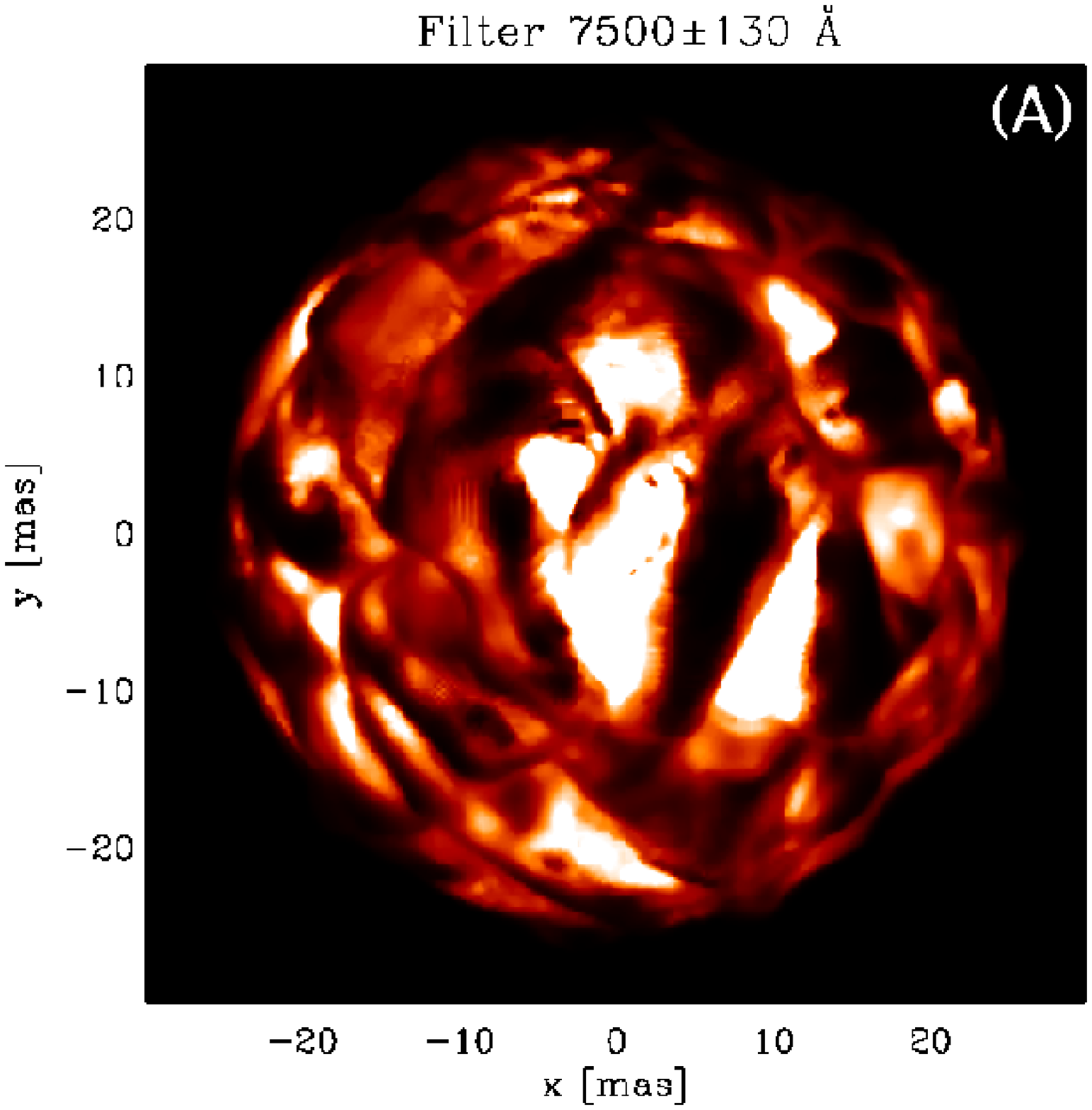}\\
\includegraphics[width=0.36\hsize]{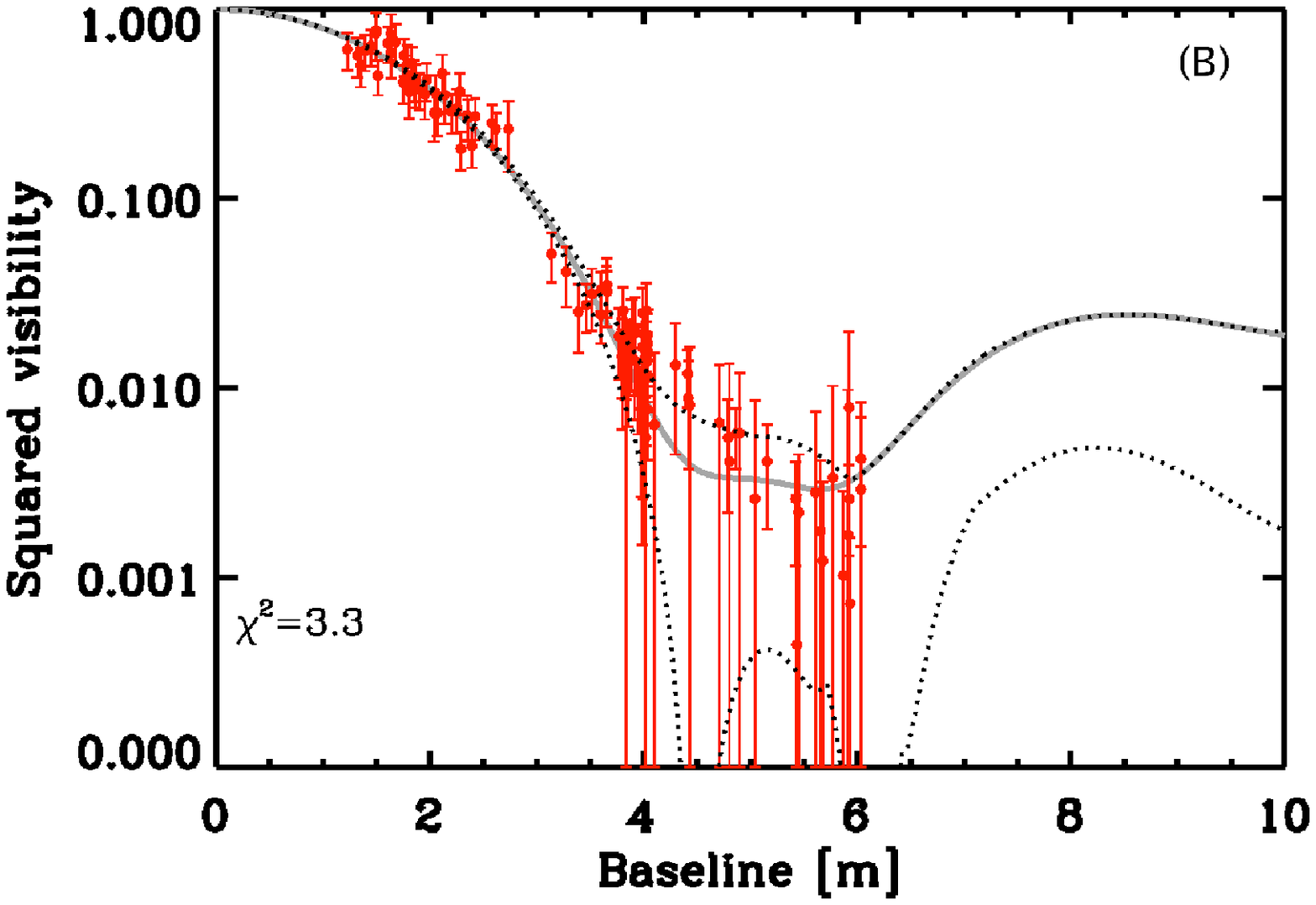}%COAST5
        \includegraphics[width=0.36\hsize]{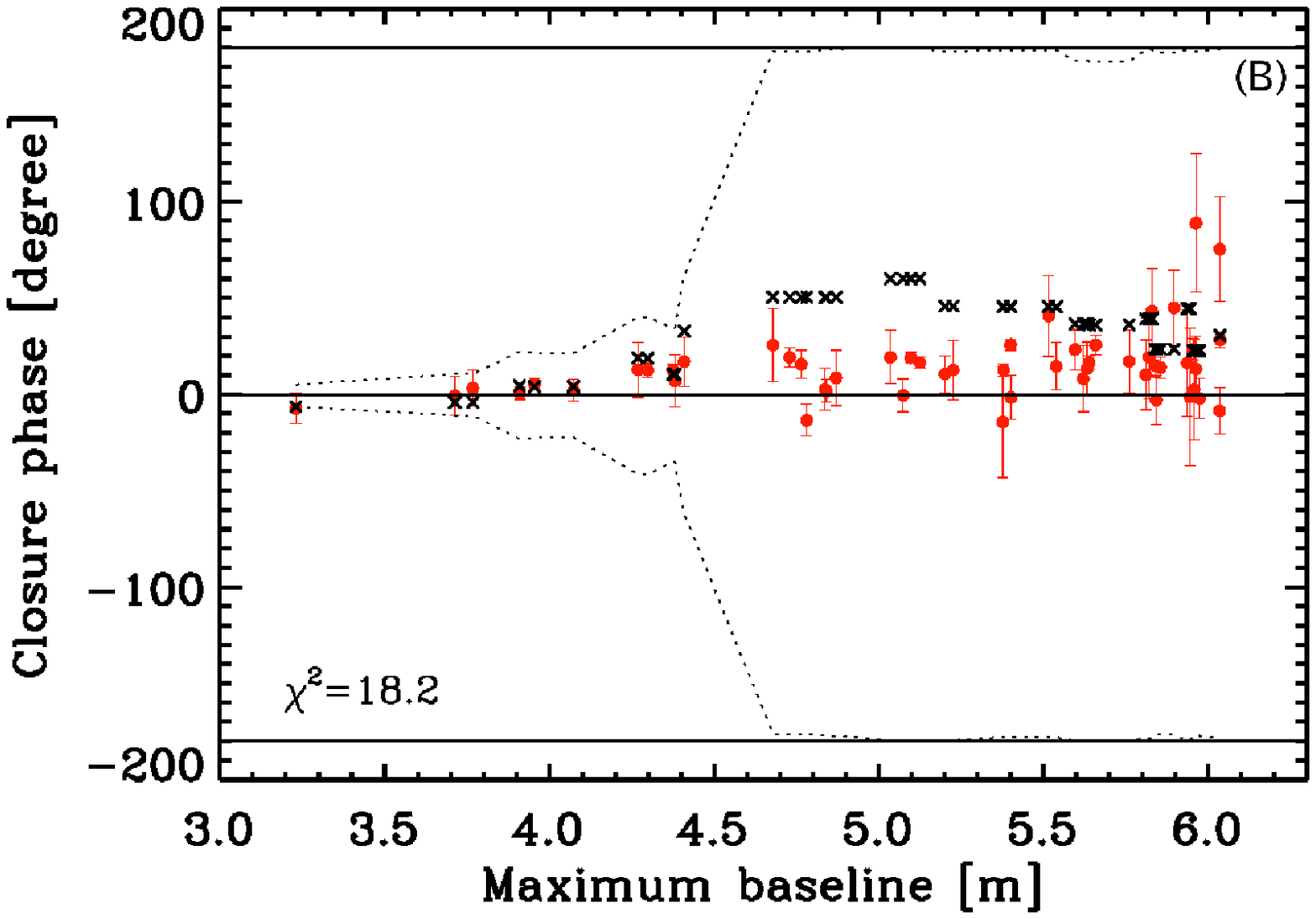}
        \includegraphics[width=0.27\hsize]{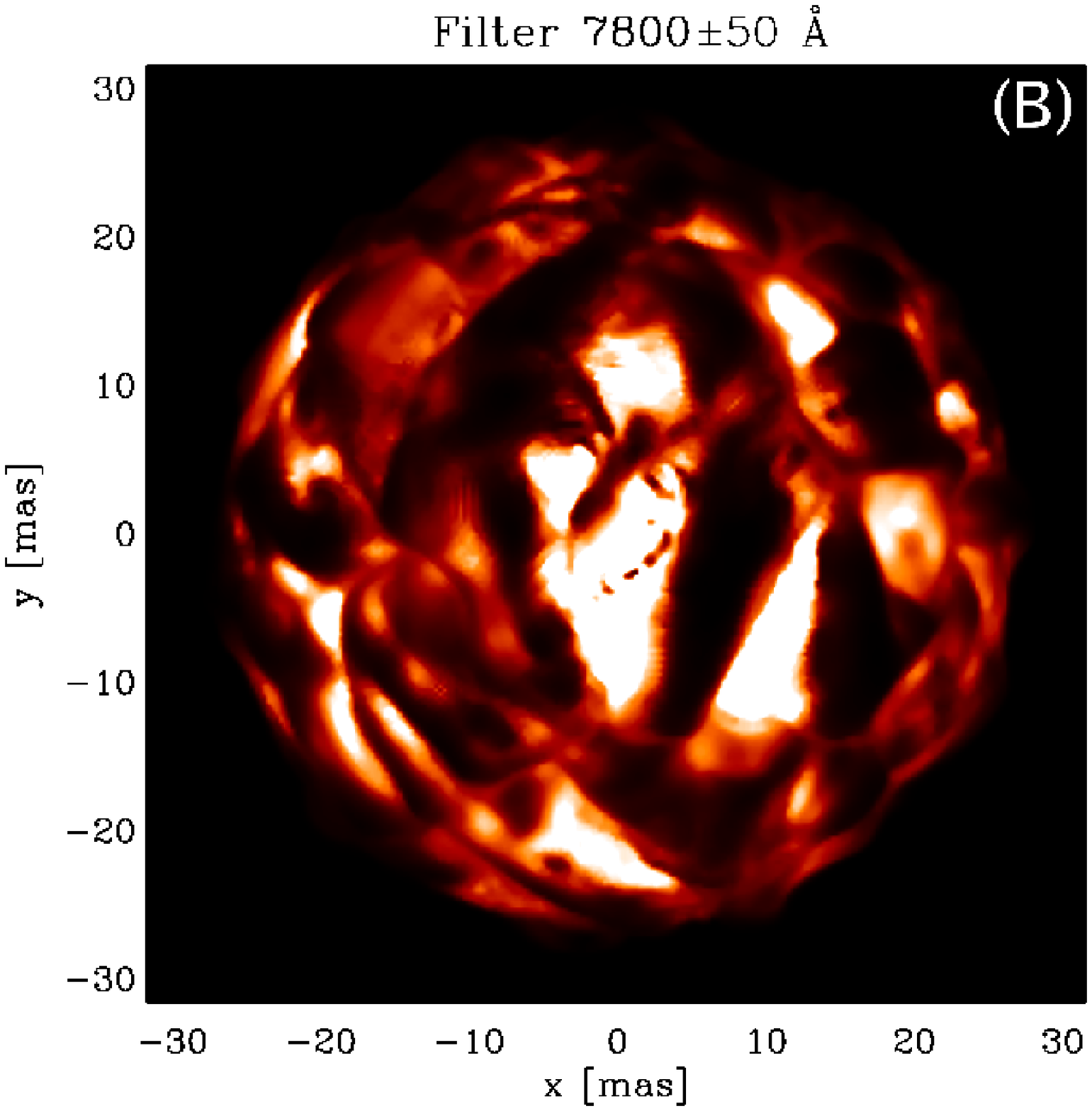}\\
\includegraphics[width=0.36\hsize]{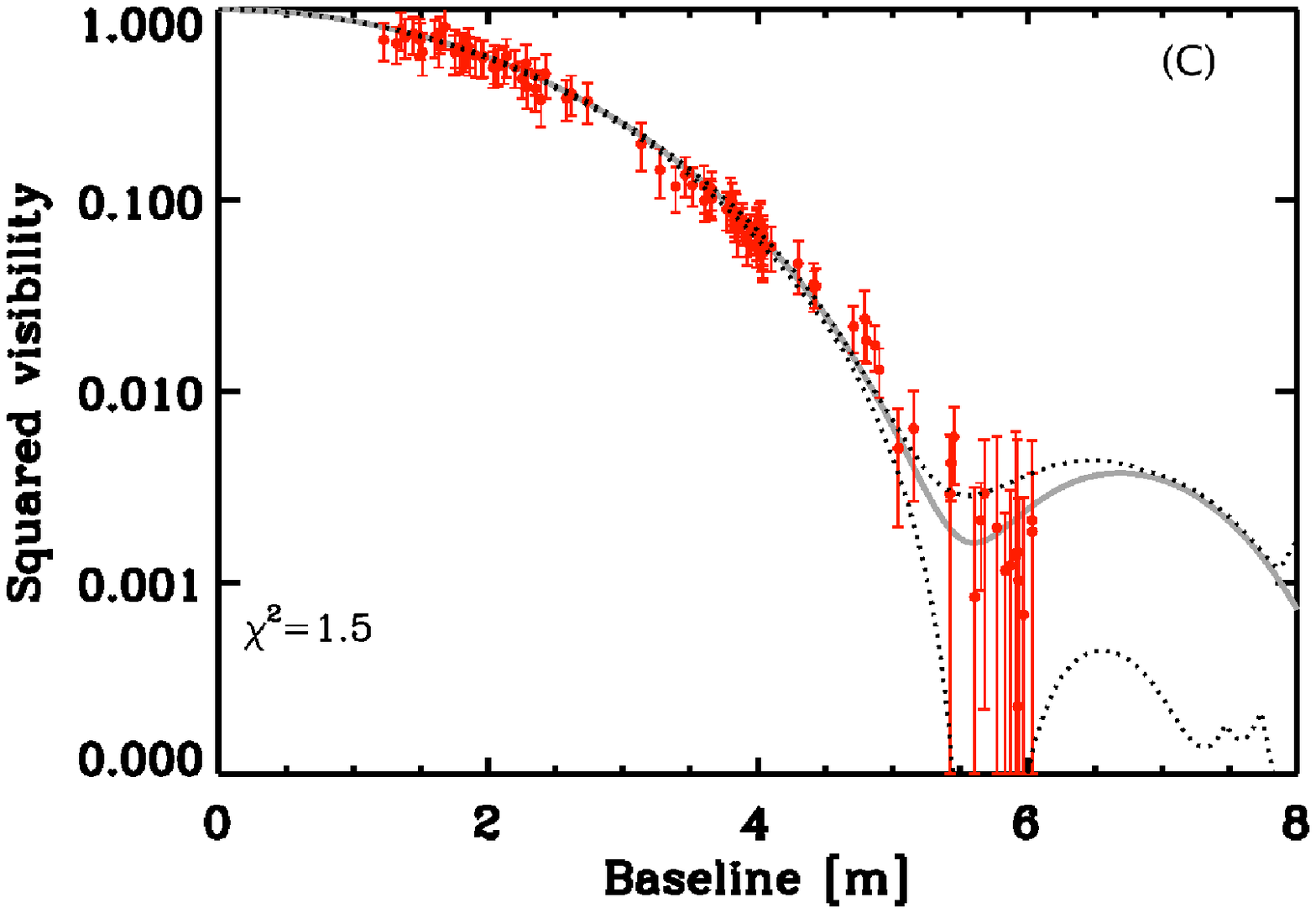}%COAST2
	\includegraphics[width=0.36\hsize]{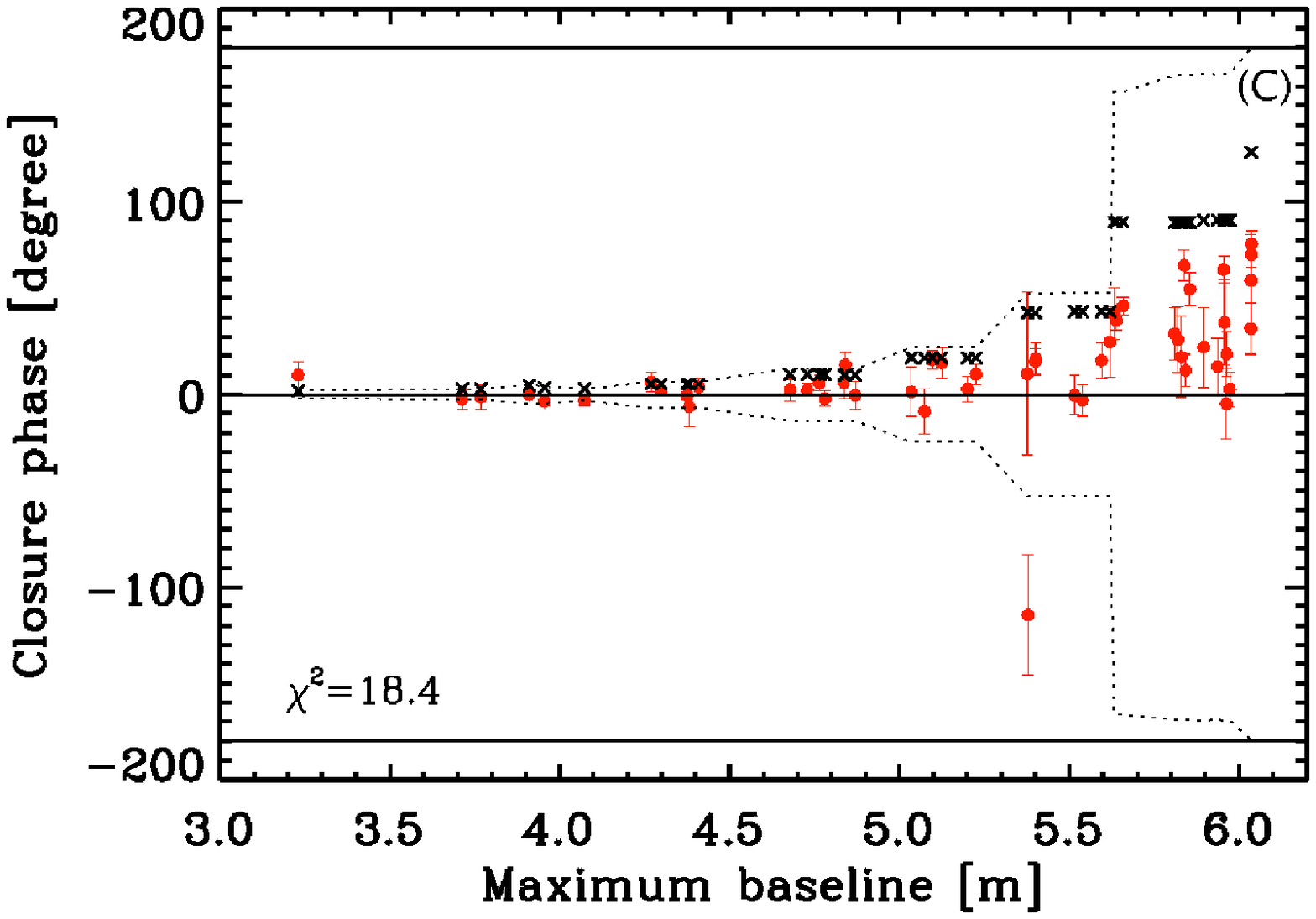}
        \includegraphics[width=0.27\hsize]{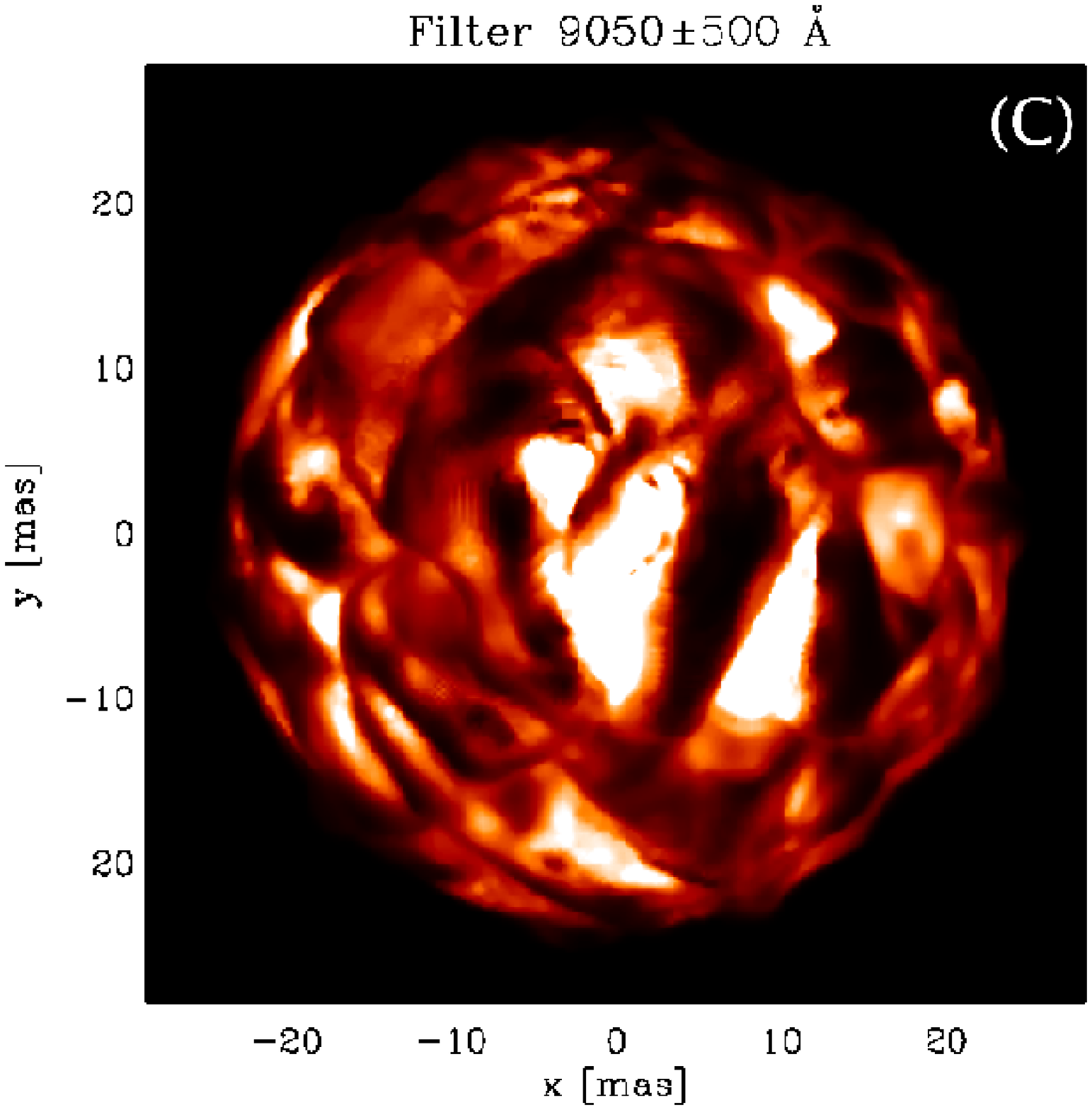}\\
\end{tabular}
        \caption{Same as in Fig.~\ref{young_comparison1} but for the
          observations of \cite{2004young}. The stellar parameters of
          this snapshot are: $L = 93\,800\,L_\odot$, $R =
          840.3\,R_\odot$, $T_{\rm eff} = 3484$\,K and $\log(g) =
          -0.35$.  The row marked with (A),
            (B) and (C) correspond to bandpass centered at 7500, 7820
            and 9050 \AA , respectively. The intensity map centered at $7500$ \AA \ has been
          scaled to an apparent diameter of $\sim$49.7 mas, at a
          distance of 157.9 pc; the map at $7820$ \AA \ to an apparent
          diameter of $\sim$47.3 mas, at a distance of 165.7 pc; and
          the map $9050$ to an apparent diameter of $\sim$52.6 mas, at
          a distance of 149.1 pc.}
         \label{young_comparison2}
   \end{figure*}

Fig.~\ref{young_comparison2} shows the comparison to the data taken in
2004. There, the filters used span wavelength regions corresponding to
TiO absorption bands with different strengths centered at
7500\AA\, 7800\AA\, and 9050\AA\ (transitions ${\rm A}^3\Phi - {\rm  X}^3\Delta (\gamma)$ and ${\rm E}^3\Pi - {\rm  X}^3\Delta (\epsilon)$, see Fig.~\ref{filters}). Again, the same snapshot fitted the whole dataset
from the same epoch.

The departure from circular symmetry is more evident than in the H
band, with the first and the second lobes already showing large visibility
fluctuations.  The RHD simulation shows an excellent agreement with
the data both in the visibility curves and closure phases.

% [JSY: the following is confusing and repeats what is said elsewhere:]
% As a consistency check, we managed to fit the observations not only using
%the same rotation angle for the visibility curves and closure phases
% but also the same snapshot for the data taken in the same epoch.

However, within the same observation epoch we had to scale the size of
the simulated star to a different apparent diameter at each observed
wavelength. For example, in Fig.~\ref{young_comparison2} the apparent
diameter varies from 47.3 to 52.6 mas. Thus we infer that our RHD
simulation fails to reproduce the TiO molecular band strengths probed
by the three filters. As already pointed out in Paper~I, our RHD
simulations are constrained by execution time and they use a grey
approximation for the radiative transfer that is well justified in the
stellar interior and is a crude approximation in the optically thin
layers; as a consequence the thermal gradient is too shallow and weakens
the contrast between strong and weak lines
\citep{2006sf2a.conf..455C}. The intensity maps look too sharp with
respect to the observations. The implementation of non-grey opacities
with five wavelength bins employed to describe the wavelength
dependence of radiation fields \citep[see][for
  details]{1994A&A...284..105L, 1982A&A...107....1N} should change the
mean temperature structure and the temperature fluctuations. The mean
thermal gradient in the outer layers, where TiO absorption has a large
effect, should increase. Moreover, in a next step, the inclusion of
the radiation pressure in the simulations should lead to a different
density/pressure structure with a less steep decline of density with
radius. We expect that the intensity maps probing TiO bands with
different strengths will eventually show larger diameter variations
due to the molecular absorption as a result of these refinements.

\begin{figure*}
   \centering
  \begin{tabular}{cc}
\includegraphics[width=0.32\hsize]{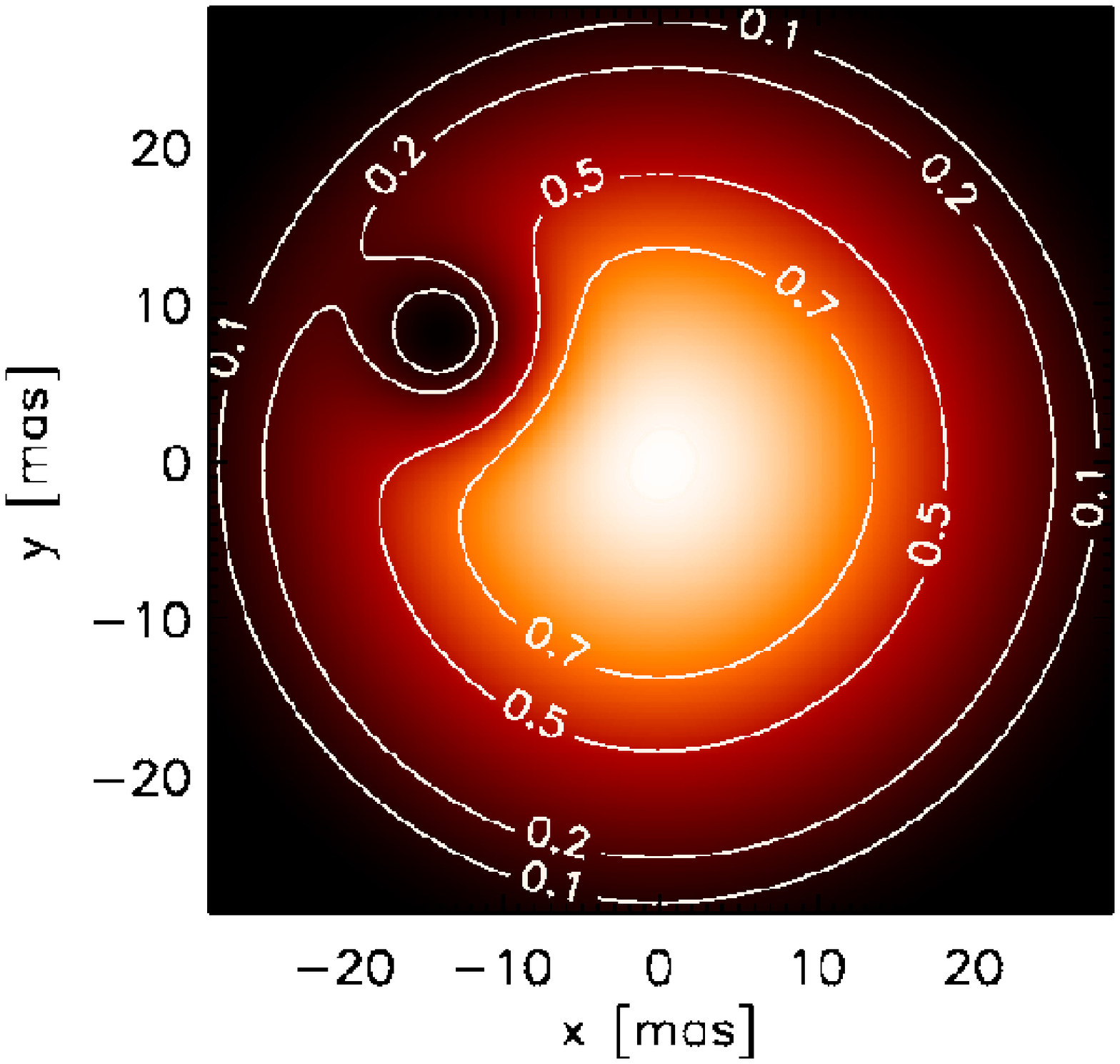}
       \includegraphics[width=0.32\hsize]{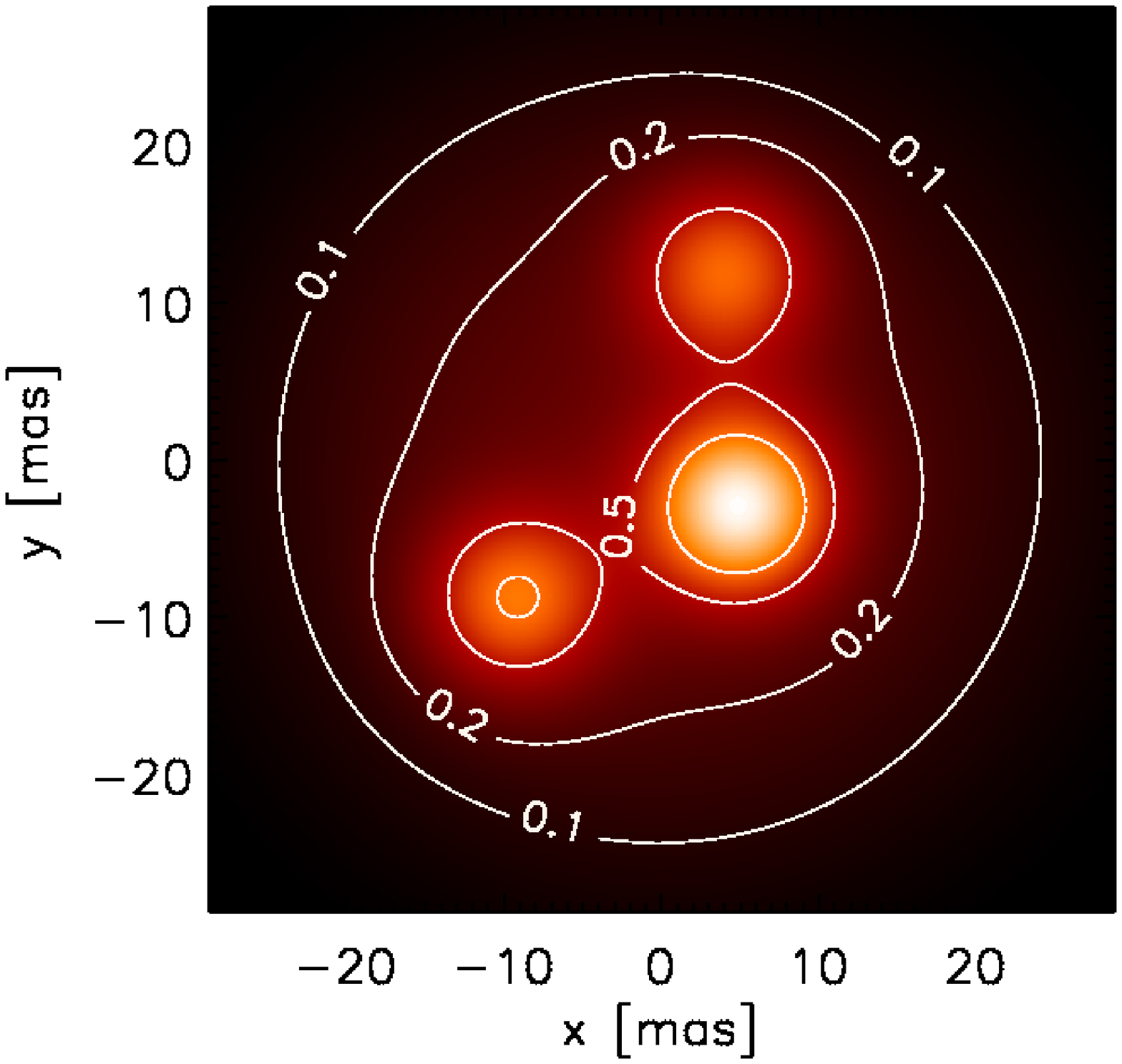}
 \includegraphics[width=0.32\hsize]{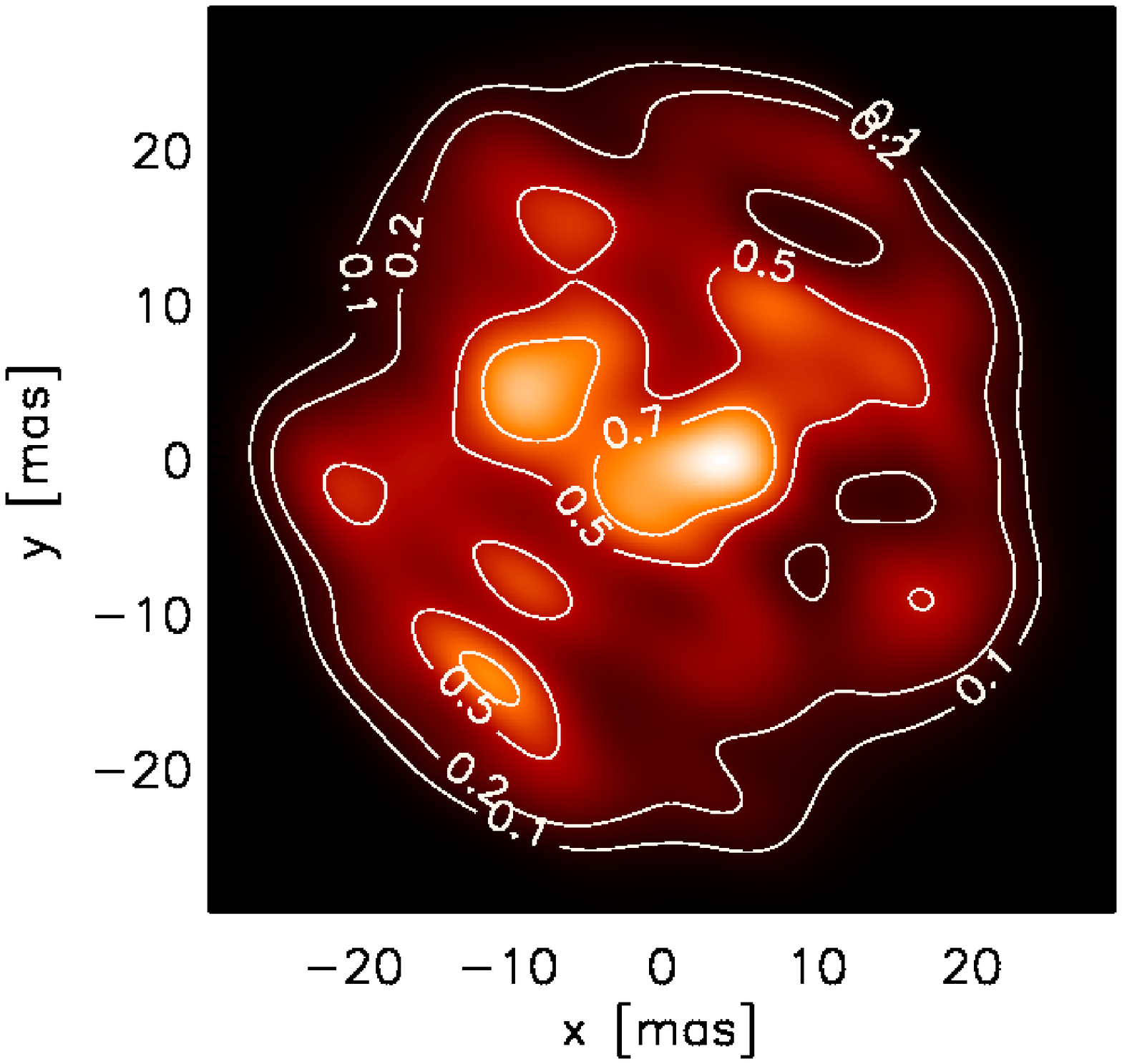}
 \end{tabular}
\caption{\emph{Left and central panels:} Brightness distributions of
  the best fitting parametric models from
  \cite{2000MNRAS.315..635Y} with dark features (left) and bright
  features (center). \emph{Right panel:} our best fitting 3D
       simulation snapshot of Fig.~\ref{young_comparison1} (row marked
       with (A)) convolved
       with a 10$\times$10 mas PSF (i.e.\ the size of the bright
  features in the parametric model).
       The intensities in both panels are normalized to the range
      [0, 1] and some contour lines are indicated
       (0.1,0.2,0.5,0.7).}
              \label{images_2}%
\end{figure*}

In \cite{2000MNRAS.315..635Y}, the authors managed to model the data
in the 7000 \AA\ filter with two best-fitting parametric models,
consisting of a circular disk with superimposed bright features
(Fig.~\ref{images_2}, central panel) or dark features (left
panel). We compared these parametric models to our best-fitting
synthetic image of Fig.~\ref{young_comparison1} (top row). The
convolved image, displayed in Fig.~\ref{images_2} (right panel), shows
a better qualitative agreement with the bright features parametric
model. In fact, 3D simulations show that the surface contrast is
enhanced by the presence of significant molecular absorbers like TiO
which contribute in layers where waves and shocks start to
dominate. The location of bright spots is then a consequence of the
underlying activity.

\section{Conclusions}

We have used radiation hydrodynamics simulations of red supergiant
stars to explain interferometric observations of Betelgeuse from the
optical to the infrared region.

The picture of the surface of Betelgeuse resulting from this work is
the following: (i) a granulation pattern is undoubtedly present on the
surface and the convection-related structures have a strong signature
in the visibility curve and closure phases at high spatial frequencies
in the H band and on the first and second lobes in the optical
region. (ii) In the H band, Betelgeuse is characterized by a
granulation pattern that is composed of convection-related structures
of different sizes. There are small to medium scale granules (5--15
mas) and a large convective cell ($\approx$30 mas). This supports
previous detections carried out with our RHD simulation in the K band
(Paper~I), using parametric models and the same dataset
\citep{2009A&A...508..923H}, \cite{2009A&A...504..115K} with VLT/NACO
observations and \cite{2009A&A...503..183O} with VLTI/AMBER
observations. Moreover, we have demonstrated that H$_2$O molecules
contribute more than CO and CN to the position of the visibility
curve's first null (and thus to the measured stellar radius) and to
the small scale surface structures. (iii) In the optical, Betelgeuse's
surface appears more complex with areas up to 50 times brighter than
the dark ones. This picture is the consequence of the underlaying
activity characterized by interactions between shock waves and
non-radial pulsations in layers where there are strong TiO molecular
bands.

These observations provide a wealth of information about both the
stars and our RHD models. The comparison with the observations in the
TiO bands allowed us to suggest which approximations must be replaced
with more realistic treatments in the simulations. New models with
wavelength resolution (i.e., non-grey opacities) are in progress and
they will be tested against these observations. From the observational
point of view, further multi-epoch observations, both in the optical
and in the infrared, are needed to assess the time variability of
convection.
%% Moreover, in order to constrain another approach a complete survey of
%% the closure phases of nearby RSG stars at lower and higher spatial
%% frequencies in order to delineate how common is the departure from
%% point symmetric brightness distribution.

\begin{acknowledgements}
This project was supported by the French Ministry of Higher Education
through an ACI (PhD fellowship of Andrea Chiavassa, postdoctoral
fellowship of Bernd Freytag, and computational resources). Present
support is ensured by a grant from ANR (ANR-06-BLAN-0105). We are also
grateful to the PNPS and CNRS for its financial support through the
years. We thank the CINES for providing some of the computational
resources necessary for this work.
\end{acknowledgements}

\bibliographystyle{aa}
\bibliography{biblio.bib}

\clearpage

\end{document}